# Predictive Capability Maturity Quantification

# using Bayesian Network

**Linyu Lin, first author**[1]
North Carolina State University
2500 Stinson Dr., Raleigh, NC, 27695
llin7@ncsu.edu

**Nam Dinh, second author**
North Carolina State University
2500 Stinson Dr., Raleigh, NC, 27695
ntdinh@ncsu.edu

**ABSTRACT**

*In nuclear engineering, modeling and simulations (M&Ss) are widely applied to support risk-informed safety analysis.*

*Since nuclear safety analysis has important implications, a convincing validation process is needed to assess simulation adequacy, i.e. the degree to which M&S tools can adequately represent the system quantities of interest. However, due to data gaps, validation becomes a decision-making process under uncertainties. Expert knowledge and judgments are required to collect, choose, characterize, and integrate evidence toward the final adequacy decision. However, in validation frameworks CSAU: Code Scaling, Applicability, and Uncertainty (NUREG/CR-5249) and EMDAP: Evaluation Model Development and Assessment Process (RG 1.203), such a decision-making process is largely implicit and obscure. When scenarios are complex, knowledge biases and unreliable judgments can be overlooked, which could increase uncertainty in the simulation adequacy result and the corresponding risks. Therefore, a framework is required to formalize the decision-making process for simulation adequacy in a practical,*

---

[1] Corresponding Author

*transparent, and consistent manner. This paper suggests a framework – "Predictive Capability Maturity Quantification using Bayesian network (PCMQBN)" – as a quantified framework for assessing simulation adequacy based on information collected from validation activities. A case study is prepared for evaluating the adequacy of a Smoothed Particle Hydrodynamic simulation in predicting the hydrodynamic forces onto static structures during an external flooding scenario. Comparing to the qualitative and implicit adequacy assessment, PCMQBN is able to improve confidence in the simulation adequacy result and to reduce expected loss in the risk-informed safety analysis.*

## 1. INTRODUCTION

Nowadays, an increasing amount of research has been conducted for developing and applying advanced modeling and simulation (M&S) tools in nuclear discipline. In risk-informed safety analysis [1] [2], M&S tools are used to investigate the effects of uncertain scenarios, simulate accident progressions, characterize the reactor safety margin, improve the operational procedures, locate design vulnerabilities, etc. Compared to classical risk analysis, the risk-informed analysis aims to address both aleatory and epistemic uncertainty within a well-defined issue space, rather than trying to work with arbitrarily defined point values of load and capacity. Meanwhile, in complex systems like Nuclear Power Plants (NPPs), since the interactions among systems, components, and external events can be highly nonlinear, risk-informed safety analysis uses advanced simulations to fully represent the generations, progressions, and interactions of accident scenarios with the NPPs. However, the classical risk-informed approach does not consider the impacts of simulation adequacy [3] [4], which includes model parameter uncertainty, model form uncertainty, numerical approximations, scaling errors, etc. As a result, a validation framework is needed to not only determine whether the M&S code is adequate for representing the issue spaces but also to be directly used in the risk-informed safety analysis.

Code Scaling, Applicability, and Uncertainty (CSAU) evaluation methodology was introduced in 1989 [5] to demonstrate a method that "can be used to quantify uncertainties as required by the best-estimate option described in the U.S. Nuclear Regulatory Commission (NRC) 1988 revision to the ECCS Rule (10 CFR 50.46) [6]". A Regulatory Guide (RG 1.203), Evaluation Model Development and Assessment Process (EMDAP), is developed at 2005 to "describe a process that the U.S. NRC considers acceptable for use in developing and assessing evaluation models that may be used to analyze transient and accident behavior that is within the design basis of a nuclear power plant [7]." In the CSAU/EMDAP framework, the complexity of physics and phenomena is emphasized, and scaling analysis is suggested to resolve the lack of data issues. The objective is to ensure both the sufficiency and necessity of validation data, modeling, and simulation, such that the simulation can adequately describe the scenarios investigated. Although the evidence involved is objective, the assessment process requires subjective information, including phenomena ranking and identification, decisions regarding data applicability, selection of validation metrics, etc. In CSAU/EMDAP, such subjective evidence is treated implicitly, and it causes the validation process to lack transparency. Meanwhile, due to a lack of formalized treatment, it becomes hard for analysts and decision makers to ensure the consistency of elicitation and processing of subjective information. Therefore, a decision model is needed for integrating all sources of evidence and determining final simulation adequacy. Meanwhile, the decision model needs to be practical, transparent, and consistent, such that the simulation adequacy results can be used with sufficient confidence.

The Predictive Capability Maturity Model (PCMM) [8] was developed by W.L. Oberkampf et al. in 2007. As a decision model for Verification, Validation, and Uncertainty Quantification (VVUQ), PCMM explicitly treats the model credibility/uncertainty assessment as a decision-making process. For designated scenarios, six attributes are designed and assessed qualitatively based on a PCMM matrix, which is designed according to the context and consequence of applications. Since the final decisions are informed by requirements and consequences, PCMM can effectively guide the development and validation of M&S tools. However, since the PCMM matrix is constructed using descriptive statements, the representations of performance standards can be obscure and suggest inconsistent criteria. Meanwhile, although validation and uncertainty quantification are discussed as major attributes, other critical components, including scaling analysis, data applicability, data quality, etc., are not explicitly discussed. As a result, when there are data gaps induced by differences between the prototypical and experimental systems, such implicitness could suggest results in inconsistent maturity levels.

Other frameworks include "Guide for Verification and Validation in Computational Solid Mechanics (VV10)" [9] and "Standard for Verification and Validation in Computational Fluid Dynamics and Heat Transfer (VV20)" [10] by ASME for quantifying the degree of accuracy to consider the errors and uncertainties in both the solution and the data. Since the adequacy results are used to support nuclear risk analysis, while VV10 and VV20 are designed as a general guidance for the V&V of computational model, CSAU/EMDAP and PCMM are more appropriate and relevant to the context of this study.

In this paper, a new decision model named Predictive Capability Maturity Quantification using Bayesian network (PCMQBN) is presented. Developed based on argumentation theory and Bayes' theorem, PCMQBN aims to formalize the decision-making for assessing simulation adequacy assessment such that the process is transparent, consistent, and improvable with new evidence. This paper is organized as follows: section 2 limits the scope of this study by introducing assumptions, conditions, and limitations of proposed framework. Section 3 formalizes the interpretation of simulation adequacy based on the nature of validation. Section 4 introduces PCMQBN, where the first two subsections describe technical basis that characterizes and integrates evidence based on the argumentation theory and Bayes' theorem; the last two subsections evaluates the behavior of this framework. Section 5 illustrates the application of PCMQBN in evaluating the adequacy of a Smoothed Particle Hydrodynamic simulation for predicting the hydrodynamic forces onto static structures during an external flooding scenario.

## 2. ASSUMPTION, CONDITION, AND LIMITATION

To properly identify the scope of this study, important conditions and assumptions are listed in Table 1. Category A aims to define the scope and application of this study; category B lists the assumptions in PCMQBN for formalizing the decision-making process in validations; category C suggests the conditions and assumptions used in case studies.

Assumption A1 limits the application to risk-informed safety analysis, and the objective is to determine the error distribution of the quantity of interest predicted by M&S. More specifically, this study focuses on situations with data gaps. As a result, to better characterize adequacy of M&Ss and to avoid unreliable expert judgments, this study aims to reduce the uncertainty in estimating the simulation adequacy and corresponding risks induced by such uncertainty. Assumption A3 mainly assumes that the code verification has been performed. The confidence on such assumption is built on the theory manual of NEUTRINO [11], together with code and solution verifications from various literature [12] [13] [14].

Assumption B1 suggests the formal methods to improve the reliability and robustness of the validation decision-making process. Formal methods have continuously proven its success in financial, computer system, etc. in reducing major losses due to unverified errors [15]. It is argued that the formal methods do not obviate the need for testing, experiments, and other assertion techniques, it is mainly designed to help identify errors in reasoning which could be overlooked or left unverified. Assumption B2 aims to formalize the validation process as an argument process and to further represent the validation argument with Bayes' theorem. However, it has been suggested that the prior probability and likelihood cannot be known precisely [16] [17]. In this study, a sensitivity analysis is suggested by performing standard Bayesian analysis with a class of prior and likelihood functions. Next, all important parameters, which have high impacts on the results, are carefully examined. If no significant discrepancy is witnessed, the result is claimed to be robust. Assumption B3 aims to suggest expected losses for representing the risks of adopting code predictions to risk-informed safety analysis. Since M&Ss are mainly used to support safety decisions and alternatives in designated applications, the corresponding "adequacy" should be defined based on the consequence of adopting the predicted QoIs. This study makes a table of synthetic monetary loss for each possible consequence, and the expected losses are calculated based on the simulation adequacy result.

Assumptions 1, 2 and 3 in category C are made to define the error distributions of QoIs in the case study based on the simulation adequacy results. It is criticized that the current assembly of simulation adequacy and model predictions is arbitrary. Therefore, the claim that the proposed framework can reduce uncertainty in simulation adequacy results is questionable [18]. However, at the initial developmental phase, it is acceptable to have a crude ensemble method for demonstration purposes. It is stressed that the parameters in the proposed framework are not fixed. As more evidence is gathered, the parameters need to be calibrated and refined. Moreover, since formal methods are designed to avoid or reduce unverified errors, it is argued that the validity of this claim should not be greatly deteriorated by assumptions for simplification purposes. Assumption 4 suggests a rational agent who prefers to have fewer expected monetary losses. It is criticized in [19] that the expected value cannot fully represent the agents' choices, where subjective and psychological impacts are neglected. It is argued that this study is at the scoping stage, and the objective is to formalize the decision-making process. In the future developmental stage, different decision analysis models can be tested and optimized for validation purposes.

## 3. SIMULATION ADEQUACY INTERPRETATION

To formalize the decision-making process in simulation-adequacy assessment, a consistent and transparent interpretation is needed for "simulation adequacy" as a theorical basis of the proposed framework. This section first reviews definitions from relevant works and identifies requirements for interpreting simulation adequacy. Next, the simulation adequacy is interpreted as a triplet set by answering three key questions. Meanwhile, examples are given for illustrating each of the three elements.

In validation, simulation adequacy is usually defined as the degree to which a simulation can adequately represent the system quantities of interest from the perspective of the intended uses [20]. In works by P. Athe et al. [21] and J.S. Kaizer et al. [22], the simulation adequacy is represented by a binary term "credibility – the determination that an object (in this particular instance, a model) can be trusted for its intended purpose." Furthermore, the concept of assurance case is adopted in the definition of credibility, and a "validation case" is developed for arguing the trustworthiness of a model to the decision maker for certain applications. This definition emphasizes the effects of expert belief and connections between simulation adequacy and the consequences of application. However, as an argument process, the assessment process heavily relies on expert opinions in claiming, reasoning, justifying, and reaching final goals. It could also become expensive to reach agreements when a group of experts with different backgrounds and knowledge is presented. Although the decision-making has been formalized in [22], it is suggested here that the process should be further quantified for better transparency and for reducing uncertainties due to inconsistent assessment results. In works by S. Mahadevan et al. [23], simulation adequacy is quantified by Probability Density Function (PDF) of model predictions or their uncertainty. The Bayes' theorem is used for either testing the hypothesis about model uncertainty or estimating the probability that model predictions represent the target phenomena. The uncertainty distribution definition emphasizes mathematical representations of uncertainty such that the reliability or probability of success of the model can be measured in a direct, objective, and quantitative manner. Although such interpretations are consistent and rigorous in mathematics, its applicability is greatly limited by assumptions in the assessment process. For example, the uncertainty distribution and likelihood function are assumed to have explicit forms and parameters [24], such that they can be determined by probabilistic inference. Moreover, these distributions are assumed to be fixed in different scenarios and applications. However, for situations with a lack of prototypical data, the uncertainty can be seriously distorted when it is propagated across different scales. Meanwhile, when there are multiple scales and phenomena involved in application scenarios, the inference of uncertainty distributions relies heavily on the quality of multi-physics and multi-scale models. If the multi-physics and multi-scale interactions are poorly captured, results from uncertainty inference can be misleading [25]. Meanwhile, these models are usually developed based on reduced-scale and separate-effect databases. Therefore, when the data are not fully applicable to the target applications due to scaling distortions, uncertainty due to model forms can hardly be characterized, and results from uncertainty inference can be further distorted. Besides, the quality of data also affects the results from uncertainty inference, however, the uncertainty inference can hardly capture its impacts without an informative prior [26]. As a result, although the simulation adequacy assessment needs to be quantified, the framework is suggested to have more flexible forms and adaptable structures than PDF-based distributions in uncertainty inference. At last, in the CSAU/EMDAP framework, simulation adequacy is defined by both accuracy information of model predictions and the applicability of the validation database [27].

Such a definition is more comprehensive and flexible since it not only considers the effect of scale gaps in assessing data applicability but also the uncertainty distributions of model predictions. In the present study, the interpretation of simulation adequacy will be made based on that from CSAU/EMDAP. Moreover, the impacts of scenarios and applications are also considered.

As a result, this study describes simulation adequacy as the degree to which M&S tools can adequately represent the system quantities of interest in the target applications. The objective is not only to determine if an M&S is good or bad but also to describe the uncertainty in the real application, especially when it is understood from non-prototypical data. In this study, simulation adequacy is suggested to be composed of three components: scenario, uncertainty/predictive capability levels, and beliefs. Note the purpose of this interpretation is not to resolve fundamental issues of uncertainty classifications through a sophisticated interpretation. Instead, this study focuses on practical resolutions for deciding simulation adequacy in complex engineering problems with a transparent and consistent framework. In the context of nuclear engineering, the term "transparent" requires a formalized interpretation and representation for simulation adequacy; the term "consistent" requires the formalization to have a mathematical basis, and allow for assumptions that cannot be violated in real applications; the term "practical" requires that the formalized simulation adequacy assessment should be easily applied to risk analysis. Eq. 1 shows a representation of simulation adequacy as a triplet set: scenarios, beliefs, and levels of uncertainty or predictive capability for M&S.

$$\text{Simulation adequacy} = \{\text{scenario}, \text{belief}, \text{predictive capability}\}$$

Eq. 1

The structure of interpretation in this study is similar to the triplet by Kaplan and Garrick [28] for probabilistic risk analysis. The definition for simulation adequacy aims to answer three questions:

1. What scenario does the M&S apply to?

In the nuclear accident and transient analysis, results of M&S are used to support system designs and risk management within a range of issue spaces. Meanwhile, since risk-informed safety analysis aims to address the scenario uncertainty, a scenario set $S = [S_1, \ldots, S_i, \ldots]$ is defined, and each element corresponds to one sampled scenario $S_i$ according to designated distributions. Therefore, the selections of computational methods and simulations naturally depend on the investigated scenario. Moreover, in scenarios with minor impacts, the reactor systems can be robust enough to withstand much higher loads than those being exerted. In this circumstance, safety decisions do not heavily rely on M&S results, requirements on model predictive capability and confidence do not need to be high. Similarly, when scenarios loads are likely to exceed system capacities and the uncertainty of M&S results could alter the safety decisions, the requirements on the predictive capability and confidence will be strict.

2. What is the predictive capability of M&S?

The "Predictive Capability" refers to the capability of M&Ss in predicting QoIs during accident and transient scenarios. As a major product of classical validation methods, the capability is quantified by errors between simulation results and observations. Such techniques, as validation metrics, statistical analysis, etc. are usually used. Meanwhile, Oberkampf et al. [20] represent the model's predictive capability by maturity levels, which are further explained by sub-attributes and descriptive terms. In this case, argumentation theory and corresponding techniques, including Goal Structure Notation (GSN), Claim, Argument, and Evidence notation (CAE), etc. are used.

3. What is the belief in the M&S predictive capability?

Due to imperfect knowledge and insufficient data, predictive capability cannot be precisely estimated, and belief is used to describe a state of knowledge regarding estimations. Although belief is represented by probability, it does not refer to the frequency or statistics in the sense that it does not represent a property of the 'real' world. Rather, belief describes our state of knowledge and discusses its effects on decisions. Table 2 shows an example of belief scales in probabilities together with their characteristics. This scale provides the definition of unreasonable model maturity levels as involving the independent combination of an end-of-spectrum condition with a condition that is expected to be outside the main body of the spectrum but cannot be positively excluded. The spectrum in this study refers to the spectrum of physics, scales, data applicability, and prediction errors. For example, when a solid-mechanistic code is applied to simulate fluid dynamics, its prediction errors for certain QoIs can occasionally be small at certain locations. However, the belief that this simulation generally has low prediction errors and high maturity should be low since the physics in solid mechanics are outside the spectrum of fluid dynamics; when experimental data for validating a simulation in kilometer-scale and multi-physics scenarios is collected from a centimeter facility that focuses on one of the involving phenomena, the belief that the experimental data are applicable to the target scenarios should be low since the scale are different and phenomena are separate. However, such reduced scale and separate-effect data cannot be positively excluded from the main body of the spectrum in target applications since the involving physics and phenomena are in the spectrum of target scenarios.

As a result, the objective is to find the belief $P_i(level_j)$, represented by probabilities, such that Eq. 2 can be satisfied for any investigated scenarios $S_i$ within the designated scenario set or issue space $S = [S_1, \ldots, S_i, \ldots]$; $P_s$ is the screening probability for beliefs in simulation's validation result, data applicability, and simulation adequacy for a given set of scenarios. It is to ensure consistent belief assignments across the entire issue space in the risk-informed safety analysis. Similar definitions can also be found in the Risk-Oriented Accident Analysis Methodology by T.G. Theofanous [2], which focuses on the scenarios spectrum and aims to distinguish unreasonable and small-probability events.

$$P_i(level_j) < P_s \text{ for all } S_i \qquad \text{Eq. 2}$$

Table 2 shows an example of screening probability assigned by expert knowledge. Examples are also provided assuming that an M&S simulation is applied to predict the generation and progression of surface waves in the flooding scenarios. Validation Result (VR) is assessed by comparing simulation predictions against validation databases, while Data Applicability (DA) is assessed by the scale of facilities, relevancy of phenomena, and quality of data.

The probability values $P_i(level_j)$ are computed from the probabilistic framework that represents a map of parameters in the causal relationships $\{d_i\}$, prior knowledge $\{\tilde{p}_i\}$, and decision parameters $\{k_i\}$:

$$P_i(level_j) = F(d_1, d_2, \ldots, \tilde{p}_1, \tilde{p}_2, \ldots, k_1, k_2, \ldots) \qquad \text{Eq. 3}$$

The prior knowledge $\{\tilde{p}_i\}$ and corresponding uncertainties are distributions and can be quantified according to the probability scale in Table 2. Causal relationships and decision parameters should not violate well-known physics and laws, and a synthetic model can be developed to support the value assignment. Meanwhile, they are assumed to be well-posed problems in the sense that they are not subject to major discontinuities and the uncertainty can be reduced to the parameter level without major modeling uncertainty. It is argued that the three questions above are sufficient for guiding validation activities and adequacy assessment. However, since simulation results are usually applied in risk analysis and safety decisions, the preferences and consequences of accepting certain simulation adequacy results need to be evaluated, especially when results contain uncertainties. Although such topics are beyond the scope, for completeness, this study briefly discusses a fourth attribute of simulation adequacy as an additional concern. Meanwhile, a synthetic model, together with a review of other sophisticated options, is included regarding the application of simulation adequacy results.

Since the simulation adequacy results are mostly applied to support safety-related decisions or alternatives, the adequacy should be judged not only based on model predictions and validation databases, it should also consider the target decisions. For example, in scenarios with severe consequences, requirements on belief and M&S's predictive capability levels should be more stringent than for those with less severe consequences. In the risk-informed analysis, the predictive capability level and belief should be higher for regions where loads distributions and capacity distributions overlap. If the adequacy result satisfies the requirements, a cost-benefit analysis is performed based on the consequence of simulation's uncertainty and risk. If the adequacy results does not satisfy the requirements or it is net beneficial for improving the predictive capability level and belief, additional iteration will be conducted to either continue developing new models, collecting new data, or updating the validation techniques. By adding risk and performance measurement results, the validation process becomes risk-informed in the sense that the acceptance criteria of simulation adequacy are informed by risks of target applications, which are caused by both model and scenario uncertainty.

## 4. PREDICTIVE CAPABILITY MATURITY QUANTIFICATION USING BAYESIAN NETWORK

To avoid expert biases and unreliable judgement within an implicit decision scheme, this study proposes a quantitative decision-making framework, named Predictive Capability Maturity Quantification using Bayesian Network (PCMQBN) to formalize the assessment of simulation adequacy. Considering the similarity between assurance case to simulation validation, the simulation adequacy assessment can be described as a "confidence argument" supported by evidence that justifies the claim that simulation provides reliable prediction in the domain of application. The evidence is collected from the validation framework and characterized mathematically such that it is consistent with the interpretation of simulation adequacy. Moreover, such an argument process can be further quantified by probabilities and maturity levels, and further represented graphically by Bayesian networks. In this framework, all evidence is integrated by probabilistic inferences and can be further represented graphically by a Bayesian network. At the same time, to integrate evidence from various sources, a synthetic decision model is suggested for determining the relative weights and conditional probabilities in Bayesian networks. Figure 2 shows the scheme for assessing and applying simulation adequacy by PCMQBN. Evidence of validation result and data applicability are firstly collected from validation activities guided by validation frameworks like CSAU/EMDAP. Sub-section 4.1 discusses in detail how evidence is collected and characterized consistently based on the maturity level assignment (4.1.1) and belief assignment (4.1.2). Next, the characterized evidence, together with decision parameters regarding the conditional dependencies among different evidence and attributes, are integrated for simulation adequacy results. Sub-section 4.2 discusses details of how evidence is integrated based on the argumentation theory and further quantified by probabilistic inference. To evaluate the sensitivity of decision parameters, sub-section 4.3 suggests a sensitivity analysis for the simulation adequacy result with the same set of evidence. At last, the simulation adequacy result is applied to safety analysis by assembling the predictions by Modeling and Simulation (M&S) and beliefs. At the same time, the parameter assignment and integration structures are subject to refinement. Sub-section 4.4 discusses different phases of simulation adequacy assessment based on qualities of each step.

## 4.1. Evidence Characterization

During a validation process like CSAU/EMDAP, different activities and materials, including validation databases, scaling analysis of experimental databases, simulation assessment results, phenomenon identification and ranking process, etc., are used to support the adequacy assessment of a simulation. To make better use of these materials, this study characterizes all related evidence based on the argumentation theory and the triplet definition for simulation adequacy. The characterization is required to be transparent, practical, and consistent. The term "transparency" requires a clear representation of evidence by mathematical forms such that the meaning and substance of evidence are maintained and visible. The term "practice" requires all related evidence to be effective for practical purposes and easily obtainable. In the context of nuclear safety analysis, the evidence should be characterized such that it can be directly used to support safety-related decisions. The term "consistency" requires the characterizations to be theoretically defendable, mathematically sound, and consistent with common knowledge and well-known rules.

There are various ways of characterizing evidence. Sun [29] categorizes evidence as direct evidence, backing evidence or counter evidence, based on its association with confidence. In the context of assurance case that focuses on safety [30], the evidence is defined as "the information that serves as the grounds and starting point of (safety) arguments, based on which the degree of truth of the claims in arguments can be established, challenged and contextualized". Furthermore, in Toulmin's argument model [31], the evidence is classified into six groups: claim, data, warrant, backing, qualifier, and rebuttal. Since validation shares many common features with the assurance case, Table 3 shows examples in simulation adequacy assessment for each element based on the Toulmin's argument model. In this study, information including simulation predictive capability, validation data, scaling results, data relevance, data uncertainty, assumptions, and conditions, are considered as evidence for assessing simulation adequacy. In addition, although indirect evidence, including process quality assurance, use history, M&S management, etc. [32], will affect the adequacy assessment for M&S, this study mainly investigates direct evidence for validation.

Since simulation adequacy is to estimate the degree that model predictions represent the real values, the errors, referred to as validation result, between model predictions and the validation data should be used to support the adequacy assessment. In some validation methods, simulation adequacy is interpreted as uncertainty distributions of model predictions [23]. However, it is argued that in nuclear applications, the difficulties and costs in collecting data under prototypical conditions are so high that only data from small-scale facilities and separate (or mixed) effect tests are practically obtainable. Therefore, the uncertainty distribution estimated by validation data on different scales can be significantly distorted. To avoid the problem of scaling distortion, it becomes necessary to evaluate the applicability of validation data to the target applications, referred to as data applicability, in addition to the validation result. As a result, the top claim of simulation adequacy is supported by sub-claims of validation result and data applicability. The validation result is to determine the errors between simulation results and the validation data, while the data applicability is to determine the applicability of validation data from reduced scales and experimental conditions in the context of applications. Next, the corresponding evidence is collected and evaluated.

The following sections discuss how evidence for validation results and data applicability are characterized. Specifically, the predictive capability is described by maturity levels, while the belief is represented by probability.

### 4.1.1. Maturity Level Assignment

There have been many researches performed to quantitatively measure the level of predictive capability for an M&S tool. Harmon and Youngblood [33]suggested a five-point maturity ranking scale based on the concept of credibility, objectivity, and sufficiency of accuracy for the intended use. Long and Nitta [34] suggested a 10-point scale by the concepts of completeness, credibility, and sufficiency of accuracy for the intended use. Pilch et al. [35] suggested a four-point scale dominated by the level of formality, the degree of risk in the decision based on the M&S effort, the importance of the decision to which the M&S effort contributes, and sufficiency of accuracy for the intended use. It is discussed by Pilch et al. that the maturity level of each element should be made based on the risk tolerance of the decision maker. NASA suggested a four-point scale based on the level of believability, formality, and credibility [36]. It was suggested by NASA that the credibility assessment should be separated from the requirements for a given application of M&S. In this study, the maturity level by W.L. Oberkampf [8] is used to represent and rank the predictive capability of M&Ss. It is believed that the maturity assessment and adequacy assessment should be dealt with independently as much as possible to reduce misunderstandings or misuse of an M&S maturity assessment. As a result, the maturity level in this study is defined by the intrinsic and fundamental attributes in the M&S validation and decision-making process. The objective is to objectively track all intellectual artifacts obtained during all related validation activities.

### 4.1.1.1. Validation Result

In this study, the "validation result" is defined as the comparisons between the model predictions and validation data. Based on the comparisons, maturity levels can be further defined by descriptive terms in Predictive Capability Maturity Model [8], value bounds from probabilistic or distance metrics, confidence interval, or hypothesis testing. The results from different validation metrics are in different ranges, and the corresponding interpretations can be distinct. Maupin *et al.* [37] has reviewed and tested a class of validation metrics with a synthetic example, it is found that the selection of metrics is problem dependent. For example, when both the experimental measurement uncertainty and model uncertainty is available, probabilistic metrics are more preferred than distance metrics. Otherwise, for results from deterministic models, the distance metrics are more appropriate. The descriptive terms are composed of two elements: model accuracy and performance standards. Performance standards are criteria for measuring "acceptability" of simulation accuracy, and they are defined according to applications and scenarios. These number are not fixed such that the upper and lower bounds can be floating in a single application, especially when multiple phenomena and databases are available. At the same time, it is suggested that the designation of value bounds should be consistent with the meaning of metrics outputs. For example, if hypothesis testing is used, higher values suggest a higher confidence level, and the corresponding level should be higher; if distance metrics are used, higher values usually suggest larger error, and the corresponding levels should be smaller.

When validation data is collected directly from the prototypical system, the validation result can directly support the argument of simulation adequacy. However, when the data is collected from reduced-scale facilities or separate effect tests, additional evidence is needed for assessing the simulation adequacy in target applications. Different from the maturity level definitions in PCMM, attributes of data applicability and scaling analysis are not included in the validation result. Rather, a separate evidence characterization, data applicability, is prepared to account for the effect of data relevance, scaling analysis, and data uncertainty. Meanwhile, the involvements of expert knowledge and judgment in selecting metrics and designing acceptance criteria are not included, and they will be discussed separately in the belief assessment.

### 4.1.1.2. Data Applicability

In addition to the levels from validation results, evidence of data applicability is also needed when the data is collected from reduced-scale facilities, Separate Effect Tests (SETs) or Integral Effect Tests (IETs), etc. The "data applicability" is defined by the similarity between validation facilities and reactor prototypical conditions. In this study, the maturity level of data applicability is characterized by a R/S/U grading system. The R/S/U is firstly developed by N. Dinh's works in 2013 [38] and has been used in [39] to evaluate the quality of validation data. The R/S/U system categorizes evidence of data applicability into three sub-attributes: [R]elevance, [S]caling, and [U]ncertainty, and each of them is designed according to their intrinsic properties. In this study, focuses have been put on extreme cases with binary grades for relevance and scaling attributes. In practical applications, intermediate grades can be introduced with higher resolutions. The relevance grade [R] is determined according to relationships of phenomenon and physics between application and reduced-scale validation databases. For example, the flow data collected from a curved tube is irrelevant to those in a straight tube since the phenomena are different; and the channel flows with $Re$ around 100 is irrelevant from those around 5000 since the dominating physics is different. The relevance grade is mostly determined by expert opinions. Phenomenon Identification and Ranking Table (PIRT) [40] and the corresponding quantitative version QPIRT [41] are strategies for identifying and ranking the relevance between validation databases and applications. The (physics) scaling grade [S] measures the degree of similarity between phenomena in the prototypical systems and reduced-scale experiments on the basis of physics scaling. At the same time, the scaling grade aims to determine if the validation databases are sufficient to justify extending the experimental model assessment results to applications. A formalized scaling analysis can be found in [5], and a recent review on scaling methodology can be found in [42]. In classical scaling analysis [27], dimensionless parameters are used for measuring the similarity between prototypical systems and reduced-scale facilities. If the dimensionless space of the validation databases covers the space of application, scaling analysis is claimed to be sufficient. Meanwhile, the database is claimed to be capable of representing behaviors and phenomena in the designated scenarios. For example, it is assumed that the lid-driven cavity flow can be sufficiently characterized by the Reynolds number ($Re$). It is also assumed that behaviors in the prototypical system can be represented by reduced-scale lid-driven cavity flow, while geometries, driven velocity, and fluid properties are different. As a result, the scaling grade for the validation databases can be decided by comparing the range of $Re$ for the reduced-scale database against the range under prototypical conditions. If the $Re$ range of validation databases covers that in prototypical systems, the scaling is graded as 1. Otherwise, scaling is graded as 0. In addition, scaling grade equals to 1 only if and only if the relevance [R] is not 0. Moreover, considering the effects of measurement errors, the uncertainty grade [U] is suggested for measuring the data uncertainty due to instrumentation errors and limited resolution.

For example, the data applicability assessment is performed when the target application is a channel flow, and the quantities of interest are the averaged flow velocity $v_0$. It is assumed that the flow can be fully characterized by Reynolds number ($Re$), and the target Re equals to $5 \times 10^3$. Meanwhile, it is required that the uncertainty, quantified by L1 relative error norm $\varepsilon_{QoI}$, in measuring $v_0$ is less than 50% of the characterized velocity $v_0$. It is further assumed that four databases are available from four different experiments. The experiment #1 is performed in a curved pipe with $Re \in [10^2, 10^4]$ and measurement error $\varepsilon_{QoI} = \pm 0.1 v_0$. The experiments #2, #3, and #4 are performed in straight pipes. Experiment #2 has $Re \in [10, 10^3]$ and $\varepsilon_{QoI} = \pm 0.1 v_0$; Both experiment #3 and #4 have $Re \in [10^2, 10^4]$, while experiment #4 has higher measurement errors $\varepsilon_{QoI} = \pm 2 v_0$. For experiment #1, since the phenomena in the curved pipe (case #1) is different from those in the straight pipe, the collected data is not relevant to the target application even though the Reynolds number and data uncertainty satisfies the target conditions. Databases of case #3 and #4 are sufficient since the physical characterization ($Re$) of validation database covers the same characterization in the target application. However, case #2 does not cover the target application. Therefore, the scaling attribute of case #3 and #4 is rated as 1, while case #2 is rated as 0. The uncertainty of case #4 in measuring quantities of interest $\varepsilon_{QoI}$ is higher than the acceptance criteria, and the corresponding attribute is rated as 0. Uncertainty of case 2 and 3 satisfies the criteria and rated to be at least 1. As a result, case #3 is found to be applicable.

### 4.1.2. Beliefs Assessment

In addition to the maturity, belief in levels of validation results and data applicability needs to be assessed based on the prior knowledge. Considering the subjective and intangible nature of beliefs, a table of belief scales is prepared for the temporary quantification of intangibles, and the results are rendered in qualitative terms by applying this scale in reverse. Table 2 provides an example with an arbitrary assignment of probabilistic values; more sophisticated evaluations might be made by different sources and groups. The objective is to reach an agreement on a single or a class of scales, and the defense in depth is assured with better scrutability and communicability [2]. Also, beliefs can be estimated by metrics, including confidence interval, probability boxes, etc. [16] [37]However, their results cannot violate Eq. 2 such that an adequate margin can be ensured. Meanwhile, the belief can also be assessed based on expert opinions and represented by splinter probabilities. The value assignment in this study is arbitrary, and it is also suggested that the values are problem-dependent. For scenarios with severe consequences and small margins, the belief assessment and the belief scales can be more stringent.

It is suggested that the attributes of data applicability and validation result are not independent. For example, it has been pointed out by [37] that the selection of validation metrics depends on the uncertainty grades. It is also suggested by [27] that the gradings for scaling and relevance are also correlated. Meanwhile, the assessment for data applicability and selection of validation metrics relies on expert opinions. Considering the objective nature of maturity level and R/S/U grades by their definitions, an evidence integration process is needed for integrating intercorrelations and dependencies among attributes, subjective and objective information to the final simulation adequacy. Although GSN provides structural representations of validation arguments, no quantitative result can be obtained. To better support risk analysis and guide model selections, additional techniques are needed to quantify evidence and to transform validation arguments into computable networks.

## 4.2. Evidence Integration

To integrate evidence in a transparent and consistent manner, many studies have employed Goal Structuring Notation (GSN) to integrate evidence to final simulation adequacy with the diagrammatic notation [43]. Based on the evidence characterization, the claim of overall simulation adequacy is supported by sub-claims of validation result and data applicability, which is further argued based on the R/S/U grade. Figure 3 depicts the network of simulation adequacy assessment by GSN [44] and defines principal components in GSN. The top objective (Goal #1) is to assess the adequacy of M&S for a designated scenario, and it is argued based on sub-claims of validation results and data applicability. Furthermore, the data applicability is argued based on three sub-claims: relevance, scaling, and uncertainty (R/S/U). The goals at bottom levels are solved by corresponding evidence and corresponding characterizations.

To quantify the validation argument with mathematical languages, this work uses probabilities and connects them with logic for quantitative reasoning. Comparing to the classical logics with rigid and binary characters, probabilistic approaches soften the constraints of Boolean logic and allow truth values to be measured on a belief scale [45]. According to Eq. 3, the belief is represented as a function of causal relationships $\{d_i\}$, prior knowledge $\{\tilde{p}_i\}$, and decision parameters $\{k_i\}$. The prior knowledge, represented by probability, has been estimated as belief and collected from the validation framework, together with the evidence of validation result and data applicability. Causal relationship includes direct and indirect dependency among all attributes. Since the dependence can be uncertain, the dependence becomes conditional to all possible states of attributes or intermediate variables. Such a process enables reasoning "by assumption" and decompose the reasoning task into a set of independent subtasks. It also allows us to use local chunks of information taken from diverse domains and fit them together to form a global interference in stages, using simple, local vector operations. Since the quantification of conditional dependency relies on conceptual relationships and expert opinions, decision models are needed for assessing conditional probabilities. A validation knowledge base is constructed by quantifying components $\{d_i\}$, $\{\tilde{p}_i\}$, and $\{k_i\}$. In addition to different evidence characterizations, PCMQBN also aims to integrate evidence from different databases, and a synthetic model is needed for assessing the conditional probabilities according to their levels in relevancy, scaling, data uncertainty, data applicability, and validation results.

For better visualizations, this study uses the Bayesian network to represent the statistical relationships between different evidence and attributes. A Bayesian Network (BN) is a directed acyclic graph (DAG) that is created by using the nodes represented by circles, arrows, and the conditional probability table. Each node defines either a discrete or a continuous random variable. An intermediate node serves as a parent as well as a child node. The nodes which have arrows directed to other nodes are parent nodes and nodes that have arrows coming from other nodes are called child nodes A node that does not have any arrow coming from another node is called as the root node, and it does not have any parent node. Arrows represent direct relationships between interconnected parent and child nodes. The conditional probability table assigned to each node describes the quantitative relationships between interconnected nodes. A BN analysis is performed based on the conditional probability as in Eq. 4 and the conditional independence assumption, i.e. $P(x, y|z) = P(x|z)P(y|z)$ if and only if $x \perp y|z$. The joint probability distributions can be described by conditional probability as:

$$P(X_1, X_2, \ldots, X_n) = \prod_{i=1}^{n} P(X_i|X_1, \ldots, X_{i-1})$$

Eq. 4

The conditional independence assumption simplifies Eq. 4 further as:

$$P(X_1, X_2, \ldots, X_n) = \prod P(X_i | \text{Parent}(X_i)) \qquad \text{Eq. 5}$$

Parent $(X_i)$ is parent nodes for $X_i$; $P(X_i | \text{Parent}(X_i))$ is the conditional probability table of $X_i$; $n$ is the number of nodes in the network. A Bayesian network can also be used as an inference tool to evaluate beliefs of events when evidence becomes available. For evidence $e$, the joint probability of all the nodes can be inferred as:

$$
\begin{aligned}
P(X_1, X_2, \ldots, X_n | e) &= \frac{P(X_1, X_2, \ldots, X_n, e)}{P(e)} \\
&= \frac{P(X_1, X_2, \ldots, X_n, e)}{\sum_{X_1, \ldots, X_n} P(X_1, X_2, \ldots, X_n, e)}
\end{aligned}
\qquad \text{Eq. 6}
$$

In this study, node $X_i$ includes Simulation Adequacy (SA), Validation Result (VR), Data Applicability (DA), Relevancy [R], Scaling [S], and Uncertainty [U], and each node is further characterized by maturity levels. Based on Eq. 4 and Eq. 5, the joint probability distributions are calculated as a product of probability distributions of each of the variable's conditional on other variables. The conditional probability table is determined based on expert knowledge in casual relationships and dependencies among different nodes. Table 4 shows an example of assigning conditional probabilities when the data applicability is assessed based on evidence from R/S/U grades. First of all, it is 0% confident that corresponding data is applicable if the phenomena and involving physics are 100% not relevant; Meanwhile, the data is applicable with 100% confidence only if the data is relevant, scaling is sufficient, and data uncertainty is acceptable with 100% confidence [27]. Second, the confidence level of having applicable data drops to 60% if the data uncertainty becomes unacceptable; the level drops to 20% if the scaling becomes insufficient; the level further drops to 5% if both scaling and uncertainty are not acceptable. These number are required to be less than 100% based on findings by D' Auria [46] such that insufficient scaling and low-quality data are expected to have negative impacts on simulation adequacy assessment. However, the values are arbitrarily assigned to quantify the relative impacts due to different root causes, and it is assumed in this study that the negative impact due to insufficient scaling is higher than that due to low-quality data. Similar techniques also apply to the conditional probabilities for simulation adequacy assessment. The simulation is 100% adequate if the data is applicable and the validation result satisfies the acceptance criteria. Moreover, it becomes 30% or less confident that the simulation is adequate if either validation result or data applicability does not satisfy the criteria.

Figure 4 shows examples of the Bayesian network with the conditional probabilistic prepared with GeNie [47]. Although the data is relevant and has good quality, the confidence for applicable validation data is 20% since the dimensionless space of validation data does not cover the space of the target application. Meanwhile, since the confidence of getting an adequate simulation given an acceptable validation and not applicable data is 0.25. the confidence for an adequate simulation is 40% even the simulation predictions have good accuracy in predicting the validation data.

In practice, since multiple databases are usually used in the validation process, the overall simulation adequacy should account for impacts from multiple nodes that represent the simulation adequacy result from each database. In this study, a synthetic integration model is designed to determine the conditional probability based on the Reactor Prototypicality Parameter (RPP) and Experimental Measurement Uncertainty (EMU).

The concept of validation cubic was first suggested in [38], and the objective is to measure how close the given test conditions are to the reactor conditions in scenario of interest to the application. The term "cubic" refers to three-dimensional and normalized space, which is filled by a body of validation evidence from validation experiments. At the same time, each "dimension" is normalized to the range of 0 to 1 such that each face has a square shape. Three dimensions include Reactor Prototypicality Parameter (RPP), system decomposition, and physics models. RPP, Reactor Prototypicality Parameter, is defined as the significance of certain evidence in supporting claims in reactor conditions. In this study, a numerical value equal to 1 stands for highly significant evidence, in the sense that the data from validation experiments are relevant, sufficient, and high-quality. 0 means insignificant evidence where the data can be irrelevant, insufficient or low-quality. System decomposition represents the separation of target scenarios into sub-phenomena and sub-physics. As a result, the validation experiments can be classified into separate or mixed effect tests, where separate phenomena and physics are investigated in different facilities. Physics models refer to the micro-scale closures, equation sets, and macro-scale effective-field model for simulating the prototypical system. Figure 5 shows an example of a validation cubic. A body of evidence $(E_1, \ldots, E_i, \ldots)$ is collected from experiments with different system decomposition, i.e. Separate Effect Test (SET), Mixed Effect Test (MET), Small-Scale Integral Effect Test (SS-IET), and Large-Scale Integral Effect Test (LS-SET). Meanwhile, each evidence $E_i$ is to develop the model and to support the validation over a range of models from sub-grid-scale models (closures) to macroscale Effective-Field Model (EFM). In this study, the RPP value is proposed to integrate the dimension of system decomposition and physics model, and it represents the relative importance of each evidence from the perspective of the physics model and system decomposition. Also, it is found that the status of evidence collection and simulation adequacy support is correlated with the filling of the Cubic's upper layer (RPP->1) across physics and system decomposition dimensions.

This study suggests a synthetic model for determining the RPP values based on the ratio of scaling parameters (Sc) in the experiments $\left[Sc_{Mod_K}\right]_{EXP}$ and in the applications $\left[Sc_{Mod_K}\right]_{APP}$:

$$RPP = \left[Sc_{Mod_K}\right]_{EXP}/\left[Sc_{Mod_K}\right]_{APP} \qquad \text{Eq. 7}$$

The $Mod_K$ represents the physical process $K$ calculated from test/experimental conditions, which is also a high-ranked physics in the application conditions. $\left[Sc_{Mod_K}\right]_{EXP}$ represents the scaling parameters of $Mod_K$ in experimental conditions, while $\left[Sc_{Mod_K}\right]_{APP}$ is the scaling parameters in the application's conditions. In fluid mechanics, $Sc_{Mod_K}$ can be quantified by dimensionless parameters, like Reynolds number and Mach number, which describe the relative magnitude of fluid and physical system characteristics, such as density, viscosity, speed of sound, flow speed, etc. To determine the conditional probability, a weight factor $\psi_{E_i}$ for each evidence $E_i$ is first calculated by Eq. 8 based on the EMU and RPP. in the validation cubic model [38].

$$\psi_{E_k} \sim m \cdot EMU_J + n \cdot RPP_{K,J} \qquad\qquad \text{Eq. 8}$$

EMU is Experimental Measurement Uncertainty that measures the uncertainty of a certain experiment, and it is determined based on the level of uncertainty characterizations of experimental measurements. A similar characterization for uncertainty levels can be found in [37]. $m$ and $n$ are grades that represent the significance of experiment $J$ and the physics $K$. The experimental significance is affected by the quality and relevance of a given experiment, while the physical significance is ranked according to the PIRT process, where highly ranked phenomena and their corresponding physics should have high a significance factor $n$. Table 5 provides an example of parameter selections and their definitions in the validation cubic decision model.

Figure 6 illustrates both 2D and 3D views of the validation cubic. To demonstrate the effects of significance factors, ranges of weight factors against the EMU values are made with three arbitrarily assigned values for $m$ and $n$. The minimum bound is obtained with RPP equals 0, while the maximum bound is obtained with RPP equals 1. It is emphasized that the current formulation is to illustrate the qualitative correlations between important decision parameters, i.e. the weight of evidence, and validation evidence, including scaling parameters, experimental VUQ qualities, etc.

After determining the weight factor $\psi_{E_i}$ for each evidence $E_i$, they are normalized to $\tilde{\psi}_{E_i}$ according to Eq. 9 and used as the conditional probabilities between overall simulation adequacy $CA$ and individual simulation adequacy from separate databases.

$$P\left(SA\big|SA_{E_i}\right) = \psi_{E_i}/\sum_{i=1}^{n}\psi_{E_i} \qquad \text{Eq. 9}$$

Considering the previous discussion on validation result and data applicability, the general standards for simulation adequacy can be identified as:

**Adequate** – For the high-rank phenomena, the accuracy in predicting the quantity of interest is acceptable. The simulation can also be confidently used in similar applications with relevant, scaling, and high-quality validation databases (high R/S/U grades or answer yes). The accuracy in predicting corresponding quantities of interest should also be acceptable.

**Inadequate** – For the high-rank phenomena, the accuracy in predicting the quantity of interest is unacceptable. The simulation cannot be confidently used in similar applications with irrelevant, insufficient, or low-quality validation databases (low R/S/U grades or answer no).

The inadequacy can be caused by reasons including unacceptable validation result, irrelevant, low-quality data insufficient validation data. In classical validations, the simulation is inadequate if one of these conditions is satisfied. In the PCMQBN framework, the simulation becomes "partial" inadequate, and the degree is defined based on beliefs in probability.

### 4.3. Sensitivity Analysis

Sensitivity analysis is the study of how the uncertainty in the output of a system can be divided and allocated to different sources of uncertainty in its inputs [48]. In Bayesian-network applications, sensitivity analysis investigates the effect of small changes in numerical parameters (prior probability, conditional probability) on the output parameters (posterior probabilities). Since the design and parameter selection of PCMQBN requires expert inputs, it is necessary to evaluate that induced uncertainty in the PCMQBN framework. A list of uncertain parameters is designed, including beliefs on the levels of evidence, conditional probability, and evidence integration structures. Next, a sensitivity analysis is performed to assess the impact of each parameter on any target nodes. In this study, an algorithm by Kjaerulff and van der Gaag [49] is used for calculating a complete set of derivatives of the posterior probability distributions over the target nodes over each of the uncertain parameters. Figure 7 shows an example of a tornado plot for the Bayesian network in Figure 4. Twelve variables are sampled, including the belief in the evidence of validation result is acceptable (VR=Yes), validation data is relevant (DA_R=Yes), validation data is sufficient for scaling (DA_S=Yes), the probability of having an adequate simulation given that the data is applicable and validation result is acceptable (SA=Yes|DA=Yes, VR=No). All parameters are sampled from 0 to 1, and the width of each bar represents the range of belief values on the target attribute (Simulation adequacy = Yes). It can be found that evidence of validation result has the most significant impact on simulation adequacy. This is reasonable since the comparison between model predictions and experimental data directly represents the simulation's degree of accuracy. The conditional dependencies of simulation adequacy on data applicability and validation result have more impacts on the target belief than other dependencies.

Sensitivity analysis is a unique feature enabled by formalizing and quantifying the decision-making process. It improves the robustness of the assessment results for simulation adequacy in the presence of uncertainty. It also helps the understanding of correlations between different attributes in the validation decision-making process such that the structure can be continuously refined. Moreover, by identifying the most sensitive attribute, simulation adequacy can be improved by collecting evidence of specific phenomena, improving the model performance for local predictions, and refining the conditional-dependency parameters. In addition, the sensitivity analysis offers a simple strategy against the imprecision issue in classical Bayesian analysis, where the uncertainty is required to be measured by a single (additive) probability, and values can be measured by a precise utility function [16]. However, such an assumption is very hard to achieve in validation since the data is too few to make precise estimates on the probability and the distribution. By performing a sensitivity study on various sources of uncertainty, the standard analysis is applied to all possible combinations of the decision including parameters, evidence, integration structure, etc. Next, a class of simulation adequacy is determined, and if the class of decisions is approximately the same, it can be claimed that a robust result is obtained. Otherwise, the range can be taken as an expression of confidence from the analysis. As a "convenient" approach against the imprecision issue, this method is also known "Robust Bayes" or "Bayesian sensitivity study" [50] [51].

### 4.4. Phase of Simulation Adequacy Assessment

To manage the progress of validation activities, PCMQBN adequacy assessment, sensitivity analysis, and applications, this study defines three phases of development for grading the quality and confidence in the simulation adequacy results based on the sources and levels of uncertainties. Table 6 defines the phases of development based on the sources and levels of uncertainties in simulation adequacy assessment by PCMQBN. At each stage, evidence needs to be collected and characterized accordingly. Meanwhile, the uncertainty in each evidence, parameter, integration structure, and the final simulation adequacy need to be evaluated. Complete documentation and review of this process mark the completion of each phase. Phase 1 is designed for initial adequacy assessment. Although the uncertainty in final adequacy is large, the objective is to agree on the evidence selection, conditional dependencies, acceptance criteria, and qualitative impacts on the target applications. Meanwhile, it serves as the foundation for phase 2. Most validation activities and decision-making efforts will be conducted in Phase 2, and the objective is to have a sufficiently adequate simulation that can support designated decisions with confidence. The quality assurance for the simulation is also required to prevent defects and issues in software products. Phase 3 involves licensing and regulatory activities, and the objective is to provide confirmatory results and define a defense-in-depth in evaluation.

To illustrate the process and help the understanding, a case study is prepared for assessing the simulation adequacy for Smoothed Particle Hydrodynamics (SPH) methods in external-flooding scenarios. A validation process has been performed and discussed in [52]. The current case study is at the scoping stage, and the decision parameters are subject to sensitivity analysis.

## 5.    ADEQUACY ASSESSMENT FOR SPH METHODS BY PCMQBN

To demonstrate the capability of PCMQBN in assessing the adequacy of simulation results, this study assesses the adequacy of SPH simulations in predicting the impact forces during an external-flooding scenario. Evidence is collected from the CSAU/EMDAP framework, which is performed and explained in detail by a separate work [53]. Sub-section 5.1 describes the assessment process for simulation adequacy. Sub-section 5.2 evaluates the sensitivity of simulation-adequacy results by sampling decision parameters and evidence characterizations. Sub-section 5.3 describes the application of simulation adequacy from PCMQBN results.

There are different types of flooding scenarios evaluated by the nuclear industry, and each may have multiple criteria for adequacy acceptance. For this external-flooding example, the analysis purpose is to assess if the simulation adequacy of SPH to model impact forces when simulating the scenario of "floods damage the building structures, enter the room, and cause diesel generator (DG) malfunctioning" is acceptable. The validation framework CSAU and its regulatory guide EMDAP is used for qualitative adequacy assessment. Figure 8 shows the scheme of the CSAU-guide validation process, and results from all activities lead to a qualitative decision of simulation adequacy for SPH methods in designated applications. The SPH methods and the simulation code, Neutrino, are explained in [53].

The corresponding QoIs include the response time and the structural loads on Systems, Structures, and Components (SSCs) by floods. The response time is the time for the external floods to reach the DG building and to potentially fail the DGs, while the structural loads are the pressure forces acting on the nuclear SSCs by the floods. This study focuses on the adequacy assessment of SPH methods in predicting the structural loads. An SPH-based software, Neutrino [11], is used to simulate an external-flooding scenario.

A PIRT process is performed to rank the importance of separate phenomena for evaluating the simulation adequacy in the designated scenarios. To estimate the structural loads with sufficient accuracy, the adequacy of SPH methods in simulating the hydrodynamic forces on stationary structures is highly important. As a result, a validation database is constructed with a list of numerical benchmarks, and evidence of simulation accuracy (validation result) is collected by comparing simulation predictions against measurements from each benchmark. At the same time, a scaling analysis is performed to evaluate the applicability of all databases. Table 7 shows a list of benchmarks together with qualitative results for each assessment. In both benchmarks, the peak pressure forces are selected as the quantity of interest, and SPH simulations are performed with different simulation parameters for complete uncertainty quantification. Next, simulation results are compared against the experimental measurements, and an L1 metric (L1 relative error norm) described in Eq. 10 is used to evaluate the accuracy of SPH's performance. The accuracy is acceptable if $L_1$ is less than 20%.

$$L_1 = |\frac{QoI_{preds} - QoI_{meas}}{QoI_{meas}}|$$

Eq. 10

where $QoI_{preds}$ represents the predicted quantity of interest by Neutrino, while $QoI_{meas}$ represents the measurements from experiments. More details about the accident scenario, PIRT process, performance measurement standards, accuracy and scaling analysis can be found in [52].

It is found from the dam break benchmark that the SPH method is able to adequately predict the hydrodynamic forces on the stational object with acceptable accuracy and applicable databases. At the same time, an opposite conclusion is obtained from the moving solids in fluid benchmark since the experimental scale is too small to cover the application scenarios. Therefore, based on the collected databases, it is hard to decide whether SPH methods can predict the hydrodynamic force on solid objects with acceptable accuracy since claims from two benchmarks seem to be contradictory. To reduce uncertainty, PCMQBN is applied to assess the simulation adequacy with the validation cubic model.

## 5.1.  PCMQBN Adequacy Assessment

Since evidence from two experimental databases is used, the weight factor needs to be calculated, and Table 8 shows the assignment of decision parameters based on validation activities from CSAU/EMDAP. Parameter $m$ represents the significance of dam-break and moving-solid-in-fluid experiments. It ranges from 0 to 3, and it is mainly determined by the quality of experiment and collected data. Since the dam break data is collected by extracting graphical points from literatures, its experimental significance is rated as low (=1). The moving solid data is collected directly from experimental facilities, and repeated runs are performed to quantify the experimental uncertainties from sensors, equipment, operating conditions, etc. Therefore, the moving-solid experiment is rated as high (=3). Parameter $n$ represents the significance of physics in two experiments, and it is rated according to the PIRT process. Since both experiments are investigating the phenomenon of hydrodynamic forces on stationary structures, they are rated as high, and the corresponding value is 3. $[Sc_{Mod_K}]_{EXP}$ and $[Sc_{Mod_K}]_{APP}$ are scaling parameters in experimental and prototypical conditions respectively. A scaling analysis has been performed and discussed in [52]. A dimensionless number $x^*$ is suggested for the dam break benchmark according to Eq. 11. $L$ is the distance between the gate and the solid object, $h$ is the initial depth of surface wave.

$$x^* = h/L \qquad \text{Eq. 11}$$

For the moving-solid benchmark, the scaling analysis shows that the accuracy in predicting the buoyance force depends on the particle intensity around the solid object. Therefore, for the moving object calculation, the cube density ratio ($\rho^*$ defined in Eq. 12) and the ratio between cube volume and average particle volume ($V^*$ defined in Eq. 13), are selected as the dimensionless parameters. $\bar{V}_{dp}$ is the average particle volume defined by Eq. 14 [54], and $d$ is the initial particle diameter.

$$\rho^* = \rho_{cube}/\rho_{fluid} \qquad \text{Eq. 12}$$

$$V^* = V_{cube}/\bar{V}_{dp} \qquad \text{Eq. 13}$$

$$\bar{V}_{dp} = d^3 \qquad \text{Eq. 14}$$

Based on the scaling parameters, the RPP can be determined according to Eq. 7. The dam break has RPP equal to 1 since the range of dimensionless parameters in validation databases covers those in the application scenario. The EMU is rated according to the characterization of experimental uncertainties (Table 5). Since the dam break data does not have any uncertainty information, the uncertainty level is rated as level 1 (EMU=0.1). The uncertainty of moving solid measurements is quantified by repeated runs and rated as level 2 (EMU=0.01). At last, all parameters are substituted into Eq. 8, and the weight factor $\psi_{E_i}$ for each benchmark can be determined. They are further normalized to $\overline{\psi_{E_i}}$ such that they can be further used in PCMQBN for calculating the conditional probabilities.

Figure 9 shows the Bayesian network for simulation adequacy assessment based on evidence and decision parameters for two numerical benchmarks. It is found that the belief level on the claim that the SPH method is adequate in predicting the hydrodynamic force is 100% when the simulation adequacy is estimated solely by evidence from the dam break benchmark. This finding is consistent with the qualitative result given the simulation accuracy and data applicability for the dam break benchmark. Meanwhile, the belief level on the same claim becomes 36% when the simulation adequacy is estimated by evidence from the moving-solid experiment. This result is similar to the qualitative results where the simulation is not adequate in simulating pressures in the moving-solid benchmarks. Furthermore, it is found that the belief level for an adequate SPH simulation is 83% when evidence from both benchmarks are used. Compared to the qualitative results, there is higher confidence that the SPH simulation is adequate for the designated purposes based on available evidence. Also, the uncertainty of simulation adequacy is less than that from the qualitative assessment since the contradictory results suggest a non-informative adequacy distribution.

## 5.2. Sensitivity Analysis

Considering the uncertainty of assigning decision parameters, a sensitivity study is performed by sampling all conditional probabilities by 10% of their current values. Figure 10 shows the sensitivity tornado, and it turns out that the relative importance of two validation databases, i.e. P(SA=Adequate|SA_DAM = Adequate, SA_Moving = Inadequate), has the highest impact on the final simulation adequacy. When the conditional probability is sampled from 0.438 to 1 (currently at 0.73 based on the RPP model), the probability of having an adequate simulation ranges from 0.64 to 1.

At the scoping stage, since the evidence of scaling grade can be unverified. Another sensitivity analysis is performed by excluding the evidence on scaling grade and setting the belief in sufficient/insufficient scaling as 50%/50%. The belief in an adequate simulation has reduced to 71%. Figure 10 shows the sensitivity tornado of simulation adequacy result with uncertain scaling grade and conditional probability. It is found that the scaling analysis for the dam break benchmark has the highest impact on the simulation adequacy result. When the parameter is sampled from 0.3 to 0.7 (currently at 0.5), the simulation adequacy ranges from 0.63 to 0.8. Meanwhile, the relative importance of two validation databases, i.e. P(SA=Adequate|SA_DAM = Adequate, SA_Moving = Inadequate), still has a high impact. Therefore, it is recommended that the sufficiency analysis (scaling grade) for validation databases needs to be verified and ensured with high confidence levels. At the same time, the relative weights of two evidence from the RPP model need to be carefully examined. However, in both cases, where scaling grade and conditional probabilities are uncertain, the beliefs on an adequate simulation are higher than 50%, and it suggests a more informative distribution than the qualitative decision analysis.

## 5.3. Application of PCMQBN Adequacy Results

To further demonstrate how PCMQBN results can be used in risk-informed validation (Figure 2), a risk-informed safety analysis is performed to evaluate potential damages to SSCs of NPPs by water waves. SPH simulations are performed to determine the structural loads by a wave for 60 cycles. The cycle is defined based on the frequency of hydrodynamic pressures by the surface-wave. Figure 11 shows the predicted time transient of hydrodynamic pressure $Pr(t)$ and impulse, and 1 cycle lasts for 9.09sec. The impulse is calculated by:

$$Im(t) = \int_{T_0}^{T_0+t} Pr(t)\, dt \qquad \text{Eq. 15}$$

To evaluate damages from structural loads in each wave cycle, the pressure-impulse (P-I) diagram is calculated for each cycle. The pressure-impulse diagram is determined by finding the maximum pressure and maximum impulse in each cycle. In structural engineering, the P-I diagram is used to describe a structure's response to blast load. Depending on the P-I values in each cycle, damages to the structure by surface waves can be characterized by 4 damage levels as in Figure 12. This study uses the P-I diagrams for Reinforced Concrete (RC) structures, and the curve of damage levels is made based on experimental data from [55].

Based on the adequacy definition, the accuracy is acceptable when the L1 error in predicting hydrodynamic pressure is less than 20%. It is further assumed that when the simulation is not adequate, either due to unacceptable error or inapplicable data, the prediction will have maximally 100% L1 errors. As a result, error bands are added to the SPH predictions by:

$$y = Y_{prd} + \varepsilon_r Y_{prd} \qquad\qquad \text{Eq. 16}$$

$Y_{prd}$ is the SPH prediction for the hydrodynamic pressure and impulse, $\varepsilon_r$ is the maximum L1 error by the requirements: $\varepsilon_r$ equals to 20% when the simulation is adequate, and the accuracy is acceptable; $\varepsilon_r$ equals to 100% when the simulation is not adequate. When the simulation adequacy is uncertain, the prediction is linearly assembled based on the confidence:

$$y_{en} = P(adq) \cdot y_{adq} + (1 - P(adq)) \cdot y_{inadq} \qquad\qquad \text{Eq. 17}$$

$y_{en}$ is the ensembled predictions; $P(adq)$ is the confidence in the claim that the simulation is adequate; $y_{adq}$ is the SPH predictions with error bands when the simulation is adequate ($\varepsilon_r = 20\%$); $y_{inadq}$ is the predictions with error bands when simulation is not adequate ($\varepsilon_r = 100\%$). Figure 13 shows the distribution of P-I values onto the damage-level plots for all 60 cycles in four conditions: (1) the simulation is 100% adequate; (2) the simulation is 100% inadequate; (3) the simulation is 50% adequate and 50% inadequate; (4) the simulation is 83% adequate and 17% inadequate.

The number of cycles in each different damage levels can be found with different simulation adequacy results. Table 9 shows the number of cycles in each damage level for four distributions of P-I values based on the simulation adequacy results. If no validation decision is made, on one hand, when the simulation is presumably 100% inadequate, all damages are predicted to be severe; on the other hand, when the simulation is presumably 100% adequate, there are no severe damages, and 21 out of 60 cycles result in light damages. If validation activities are performed, and when a qualitative validation decision is made with 50/50 adequacy results, 26 out of 60 cycles (43.3%) turn to be severe. However, when a quantitative validation decision is made 83/17 adequacy results based on the PCMQBN framework, all cycles turn to be moderate.

To further demonstrate how these predictions affect the safety analysis, an expected loss $\langle C \rangle$ is calculated based on a table of synthetic monetary loss and the probability of each damage levels.

$$\langle C \rangle = \int P_{DL} \cdot C dC = \sum_{i=1}^{4} P_{DL(i)} \cdot C_{DL(i)} \qquad \text{Eq. 18}$$

$C_{DL(i)}$ is the consequence in monetary losses for the damage level $i$, and a synthetic value is assigned in Table 9; $P_{DL(i)}$ is the chances that the predicted cycles will fall into the damage level $i$, and it is determined in Table 9. $i$ ranges from 1 to 4 and it represents four damage levels from no damage to severe damage. Table 10 shows the value of expected loss $\langle C \rangle$ based on Eq. 18 and corresponding values in Table 9.

It is found that if the decision maker is willing to accept potential risks by the simulation errors and completely trust the simulation with 100% simulation adequacy, the expected loss is the smallest, which suggests an optimistic attitude to the simulation prediction errors. However, if the decision maker is not willing to accept any risks by simulation errors, the expected loss is greatest, which suggests a conservative attribute to the simulation and its prediction errors. Meanwhile, it is found that with simulation adequacy result assessed by PCMQBN (83/17), the expected loss is reduced by 30% compared to the qualitative and implicit decision framework (50/50) in classical validations. Assuming our goal is to make the expected loss less than $60. The currently available evidence is sufficient to achieve this target. However, with the qualitative decision framework, we need additional validation efforts to further improve our confidence in simulation adequacy. Therefore, it is found that compared to the qualitative decision analysis, the PCMQBN framework is able to reduce costs by effectively conducting and planning validation activities.

## 6.    CONCLUSION

In this study, a framework of PCMQBN is developed to formalize and quantify the validation decision-making process with mathematical languages. The objective is to support the decision-making process for simulation adequacy in a transparent, consistent, and improvable manner. PCMQBN first formalizes the mathematical representation of simulation adequacy as a triplet of scenario, predictive capability level, and belief. Next, argumentation theory is employed to formalize the decision-making process in validation as an argument for simulation adequacy that is based on evidence from the validation frameworks and activities. In this process, all related evidence is characterized such that its representation is consistent with the definition of simulation adequacy. Next, all evidence is quantified where the predictive capability is represented by maturity levels and the belief is quantified by probabilities. Next, Bayes' theorem is used to integrate the quantified evidence, and the Bayesian network is used to represent this integration by directed acyclic graphs. To ensure the consistency of network connections and causal dependence on well-known physics, rules, and knowledge, a synthetic model is also suggested for evaluating the conditional probability among all nodes in the network by calculating the Reactor Prototypicality Parameter. A sensitivity analysis is performed to evaluate the impact of conditional probability and decision parameters. It is found that the conditional dependency between simulation adequacy and validation result has higher impacts on those between [R]elevancy/[S]caling/[U]ncertainty grade and data applicability. It is also found that relative weights of evidence from different databases have large impacts on the final data adequacy. Therefore, during a validation decision-making process, the correlations and dependencies among different databases and attributes need to be evaluated more carefully than accuracy assessments and scaling analysis for separate models and databases. Based on the sources and levels of uncertainty, three phases of development are defined for documenting and grading the quality of the assessment process and simulation adequacy results.

To demonstrate the capability of PCMQBN, a case study is performed to assess the adequacy of SPH methods in simulating the scenario of "floods damage the building structures, enter the room, and cause diesel generator (DG) malfunctioning". The validation framework CSAU and its regulatory guide EMDAP is used for collecting evidence and qualitative adequacy assessment. Details about SPH simulations and evidence collection are discussed in [52]. Since opposite conclusions are obtained from two numerical benchmarks, the PCMQBN framework is used to further refine the adequacy assessment with quantitative results. For separate benchmarks, it is found that the belief level on the adequacy claim for the SPH method is consistent with the qualitative results from CSAU/EMDAP. Meanwhile, it is found that the belief level for an adequate SPH simulation is 83% when evidence from both benchmarks are used. Comparing to the qualitative result, there is higher confidence that the SPH simulation is adequate for the designated purposes based on available evidence. Also, the uncertainty of simulation adequacy is less than that from the qualitative assessment since the contradictory results suggest a non-informative adequacy distribution. To further demonstrate how PCMQBN results can be used in risk-informed validation, a risk-informed safety analysis is performed to evaluate potential damages to SSCs of NPPs by water waves. SPH simulations are performed to determine the structural loads by a wave for 60 cycles. Based on a synthetic ensemble model, distributions of SPH predictions and corresponding consequences are made based on the simulation adequacy results. It turns out that the expected loss determined based on the PCMQBN results is 30% less than that loss from the qualitative assessment. As a result, the formalized PCMQBN framework is able to reduce the uncertainty in simulation adequacy assessment and the expected losses in the risk-informed analysis due to that uncertainty.

**ACKNOWLEDGMENT**

This work is fully supported by the U.S. Department of Energy via the Integrated Research Project on "Development and Application of a Data-Driven Methodology for Validation of Risk-Informed Safety Margin Characterization Models" under the grant DE-NE0008530. The author would also like to acknowledge the comments and suggestions by Dr. Robert Youngblood and Mr. Steven Prescott at Idaho National Laboratory, Mr. Ram Sampath and Mr. Niels Montanari at CentroidLAB Inc., Dr. Matthieu Andre and Dr. Philippe Bardet at George Washington University.

**FUNDING**

**NOMENCLATURE**

| | |
|---|---|
| BN | Bayesian Network |
| SA | Simulation adequacy |
| CSAU | Code Scaling, Applicability, and Uncertainty |
| CFD | Computational Fluid Dynamics |
| DA | Data Applicability |
| DG | Diesel Generator |
| DL | Damage Level |
| DNS | Direct Numerical Simulation |
| EMDAP | Evaluation Model Development and Assessment Process |
| EMV | Expected Monetary Value |
| EFM | Effective-Field Model |
| GSN | Goal Structuring Notation |
| IET | Integral Effect Test |
| SS-IET | Small-Scale IET |
| LS-IET | Large-Scale IET |
| M&S | Modeling and Simulation |
| MET | Mixed Effect Test |
| NRMSE | Normalized Root Mean Squared Error |
| NPP | Nuclear Power Plan |
| NRC | Nuclear Regulatory Commission |
| PCMM | Predictive Capability Maturity Model |
| PCMQ | Predictive Capability Maturity Quantification |
| PCMQBN | PCMQ using BN |
| P-I | Pressure-Impulse |
| PIRT | Phenomenon Identification and Ranking Table |
| QoI | Quantity of Interest |
| R/S/U | Relevancy/Scaling/Uncertainty |
| RPP | Reactor Prototypicality Parameter |
| SET | Separate Effect Test |
| SPH | Smoothed Particle Hydrodynamics |

SSC     System, Structure, and Component

VVUQ    Verification, Validation, and Uncertainty Quantification

VR      Validation Result

**Figure Caption List**

Figure 1: The schematic structure for this paper.

Figure 2: Scheme for the assessment and application of simulation adequacy by PCMQBN

Figure 3: Decision model for simulation adequacy assessment in a designated scenario. Principal components and their descriptions in global structure notation (GSN) [54].

Figure 4: Example of Bayesian network for simulation adequacy assessments with designed conditional probability table by expert knowledge. The plot is prepared with GeNie.

Figure 5: Validation Cubic for a body of evidence for validating sub-grid-scale models (closures) to macroscale effective-field model EFM. Evidence included ($E1, \ldots, Ei, \ldots$) is notational and they are collected from different experiments or databases with different levels of system decompositions. The relative importance of evidence is represented by RPP values. The status of validation evidence and simulation adequacy support is correlated with filling of the Cubic's upper layer (RPP->1) across physics and system decomposition dimensions.

Figure 6: Illustration of validation cubic: left: 3D surface plot for weight factor $\psi Ei$ given $m = n = 1$; right: ranges (minimum and maximum) of weight factors $\psi Ei$ with three arbitrary values of $m$ and $n$. The uncertainty is introduced by samples of RPP values.

Figure 7: Example of tornado sensitivity plot. All listed evidence and conditional probabilities are sampled by 40% of their current values, the colored bar shows the maximally reachable ranges for the final simulation adequacy results. These ranges are arranged based on their widths, while the number indicates their ranks of importance to the simulation adequacy.

Figure 8:   Demonstration of adequacy assessment based on CSAU/EMDAP.

Figure 9:   Simulation adequacy estimated by the evidence from two benchmarks and weight factors estimated by the listed decision parameters.

Figure 10: Sensitivity plot for the simulation adequacy assessment with uncertain scaling grade and uncertain conditional probabilities. All conditional probabilities are sampled by 40% of their current values, the maximally reachable belief in an adequate simulation ranges from 0.63 to 0.8.

Figure 11: Predicted time transient of hydrodynamic pressures (left) and impulse (right) onto the structure by water surface waves.

Figure 12: Logarithm plot of damage levels. Four levels of damage are defined based on the P-I values.

Figure 13: Distribution of P-I values onto the damage levels for all 60 cycles when the simulation is 100% adequate or 100% inadequate (left) and when the simulation is 50% adequate or 83% adequate (right).

**Table Caption List**

Table 1:    Important conditions and assumptions with respect to aspects of investigation.

Table 2:    Assignment of screening probability with characteristics and examples.

Table 3:    Elements of Toulmin's Argument model with simulation adequacy example.

Table 4:    Example of conditional probabilities based on expert knowledge on causal

relationships and dependencies among different evidence characterizations.

Table 5:    Summary of parameters in the validation cubic decision model.

Table 6:    Definition of phases of development.

Table 7.    Validation results for SPH methods in simulating hydrodynamic forces on stationary

structures in the external-flooding scenario.

Table 8:    A list of decision parameters in validation cubic model. The value is assigned based

on author's knowledge.

Table 9:    Number of cycles in each damage levels for four different simulation adequacy

results.

Table 10:  Expected losses for four simulation adequacy results.

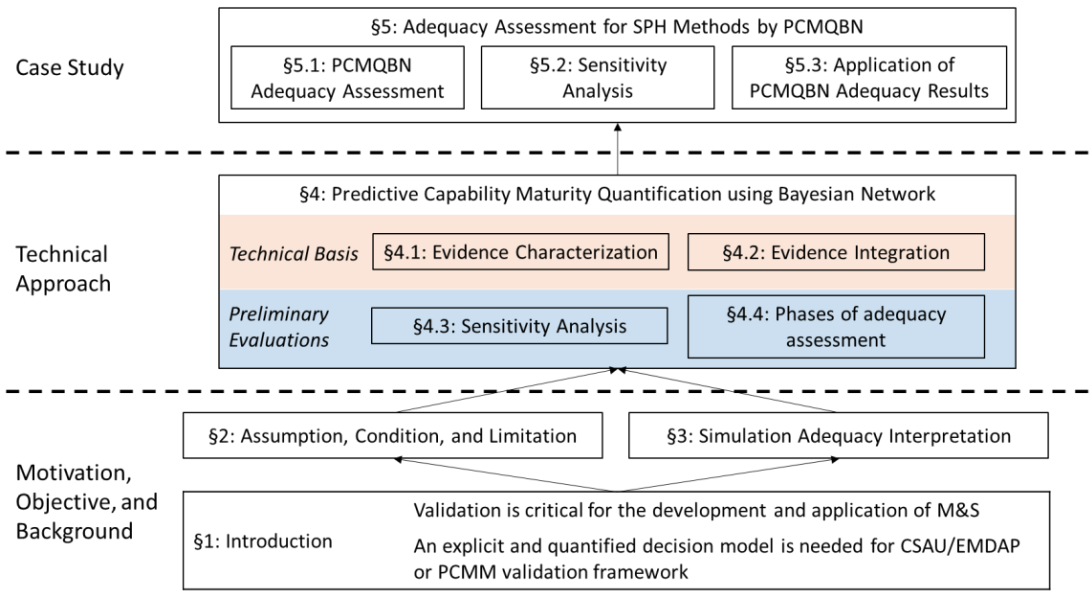

Figure 1: The schematic structure for this paper.

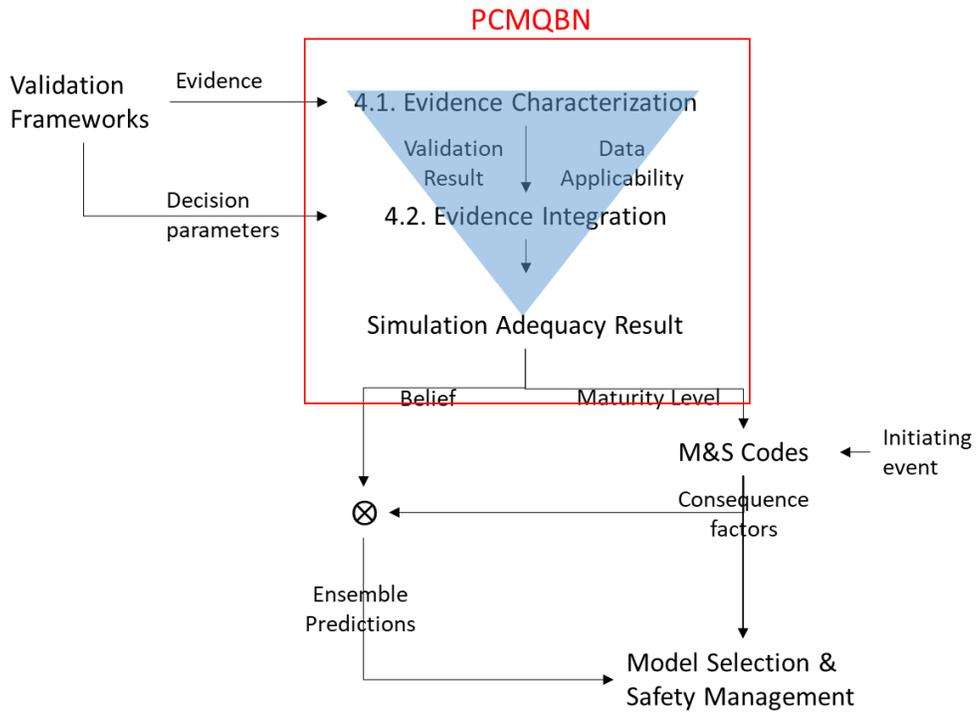

Figure 2: Scheme for the assessment and application of simulation adequacy by PCMQBN.

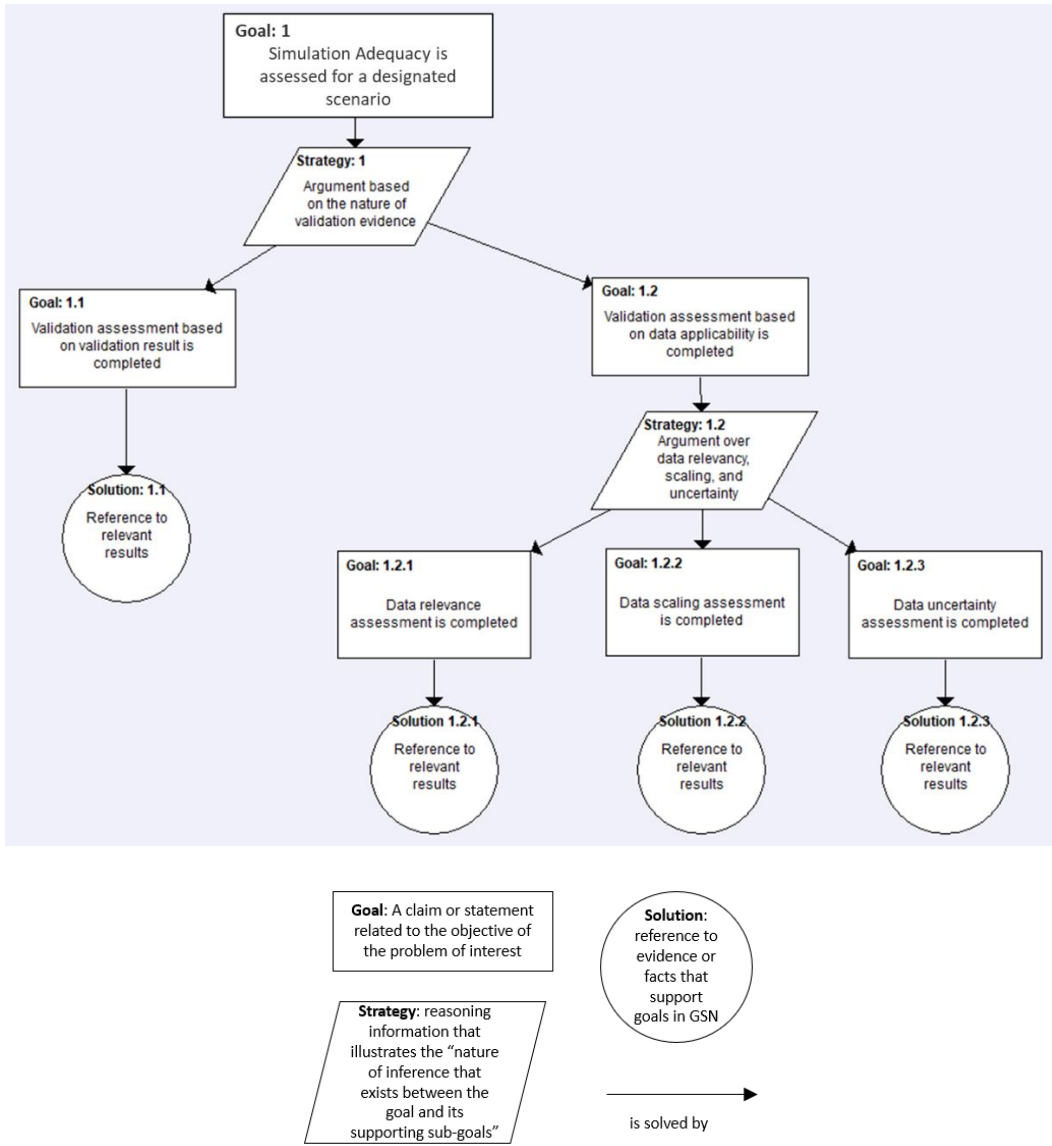

Figure 3: Decision model for simulation adequacy assessment in a designated scenario. Principal components and their descriptions in global structure notation (GSN).

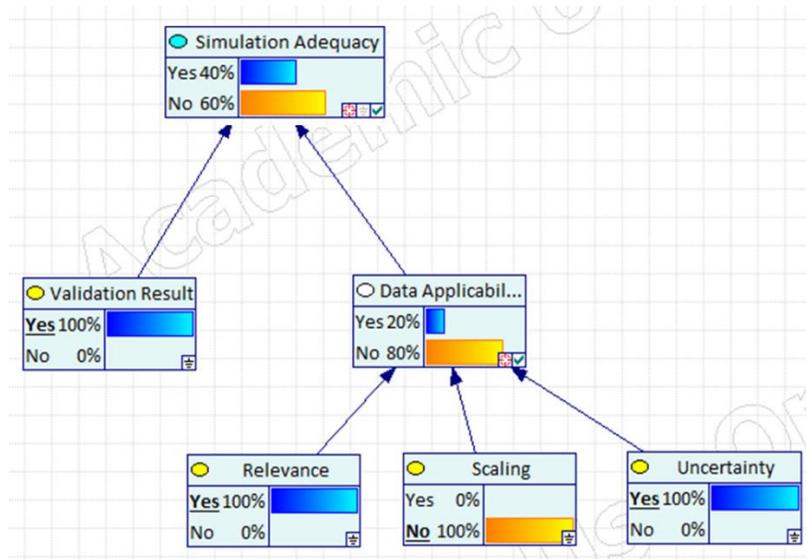

Figure 4: Example of Bayesian network for simulation adequacy assessments with designed conditional probability table by expert knowledge.

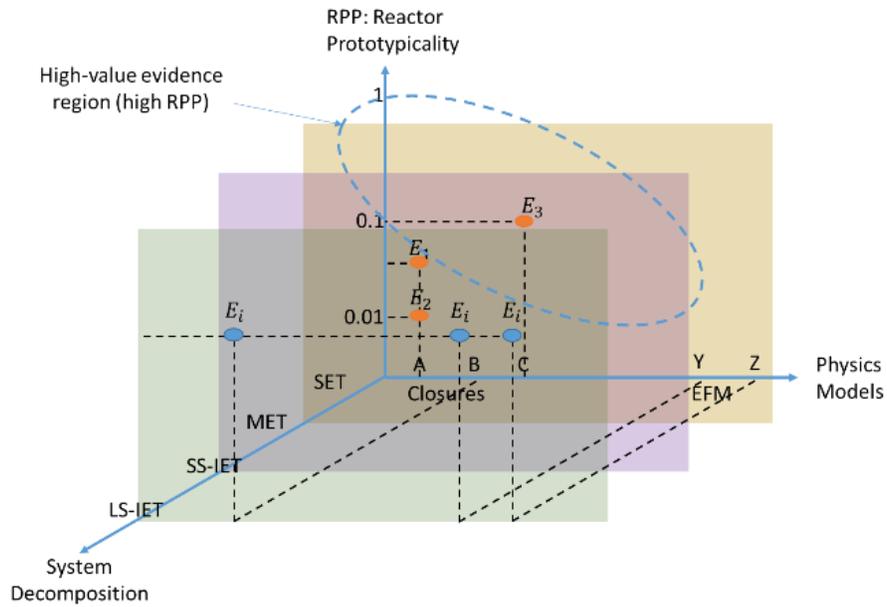

Figure 5: Validation Cubic for a body of evidence for validating sub-grid-scale models (closures) to macroscale effective-field model EFM. Evidence included $(E_1, \ldots, E_i, \ldots)$ is notational and they are collected from different experiments or databases with different levels of system decompositions. The relative importance of evidence is represented by RPP values. The status of validation evidence and simulation adequacy support is correlated with filling of the Cubic's upper layer (RPP->1) across physics and system decomposition dimensions.

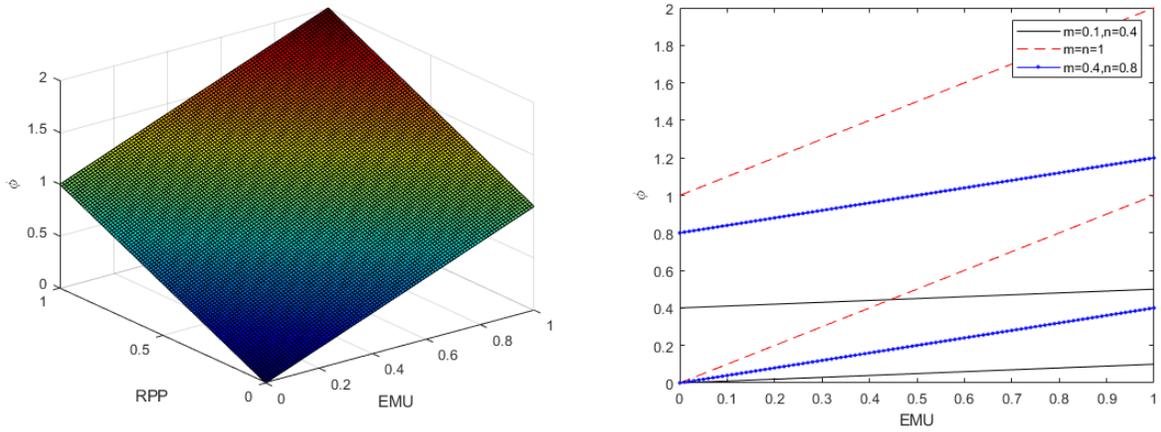

Figure 6: Illustration of validation cubic: left: 3D surface plot for weight factor $\psi_{E_i}$ given $m = n = 1$; right: ranges (minimum and maximum) of weight factors $\psi_{E_i}$ with three arbitrary values of $m$ and $n$. The uncertainty is introduced by samples of RPP values.

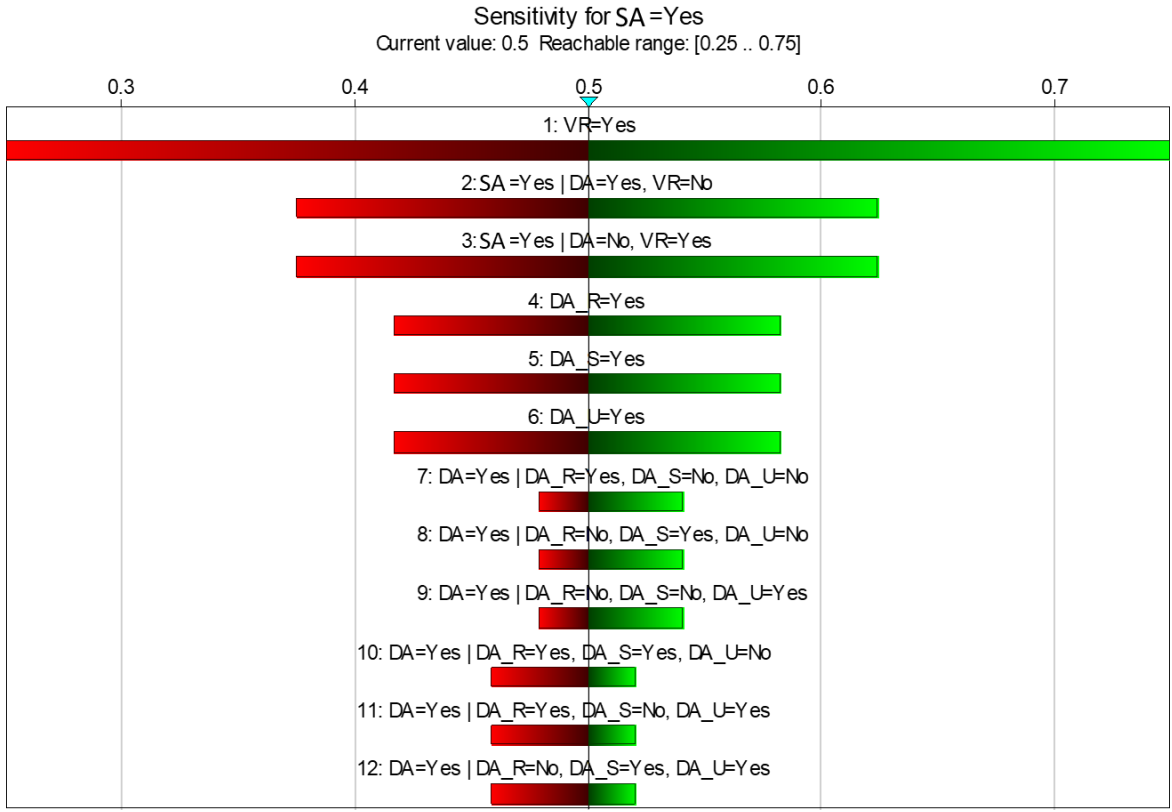

Figure 7: Example of tornado sensitivity plot. All listed evidence and conditional probabilities are sampled by 40% of their current values, the colored bar shows the maximally reachable ranges for the final simulation adequacy results. These ranges are arranged based on their widths, while the number indicates their ranks of importance to the simulation adequacy.

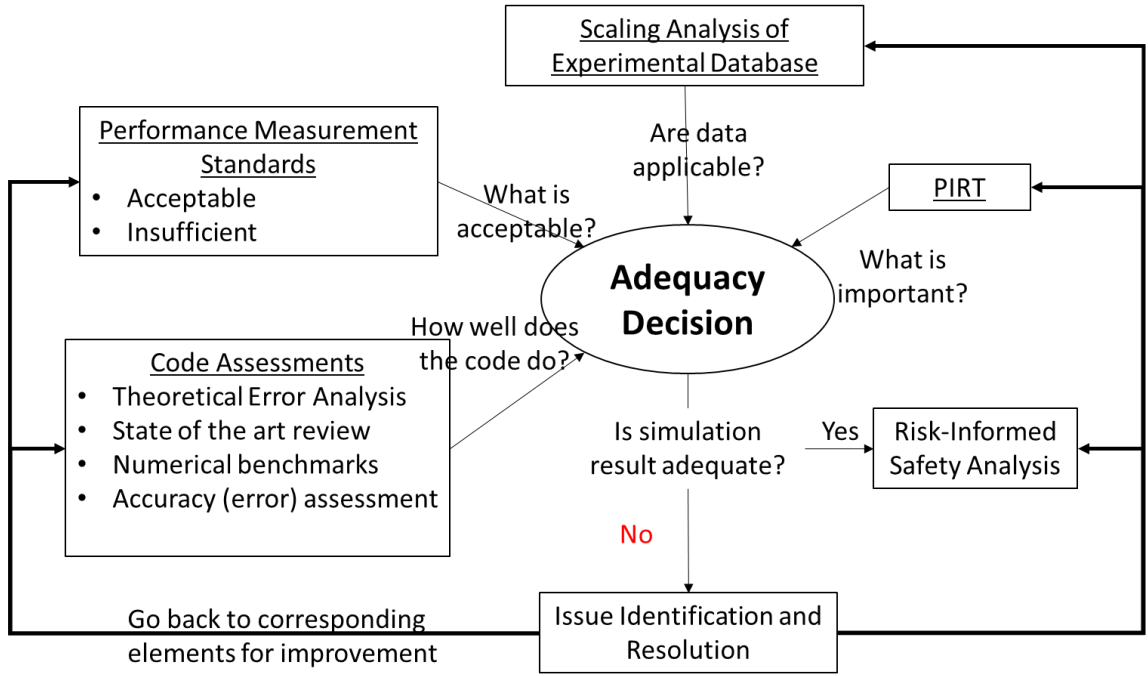

Figure 8: Demonstration of adequacy assessment based on CSAU/EMDAP.

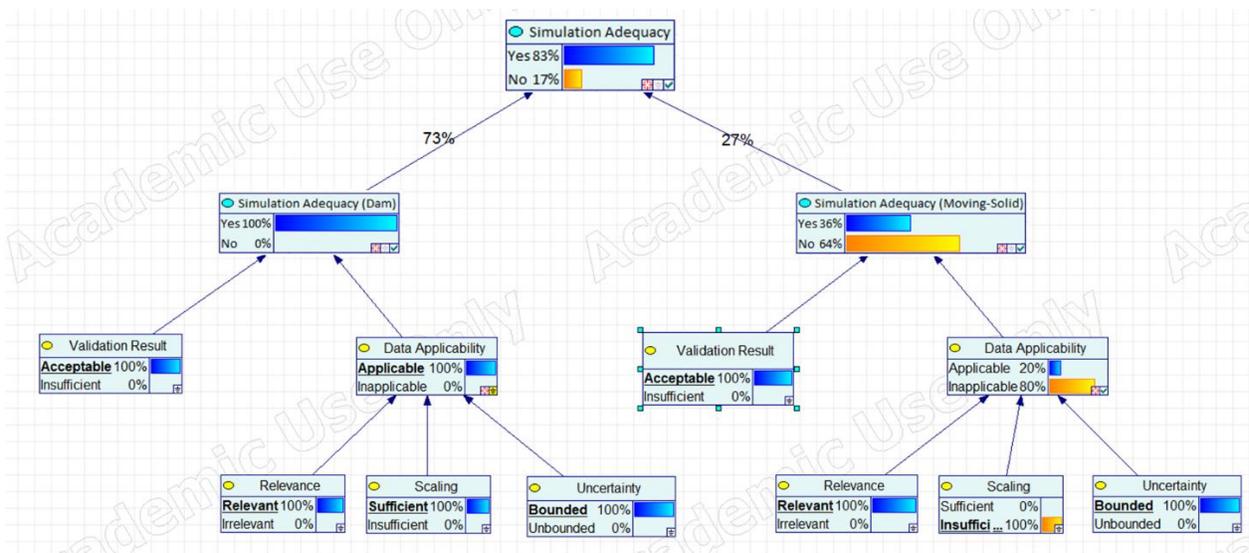

Figure 9: Simulation adequacy estimated by the evidence from two benchmarks and weight factors estimated by the listed decision parameters.

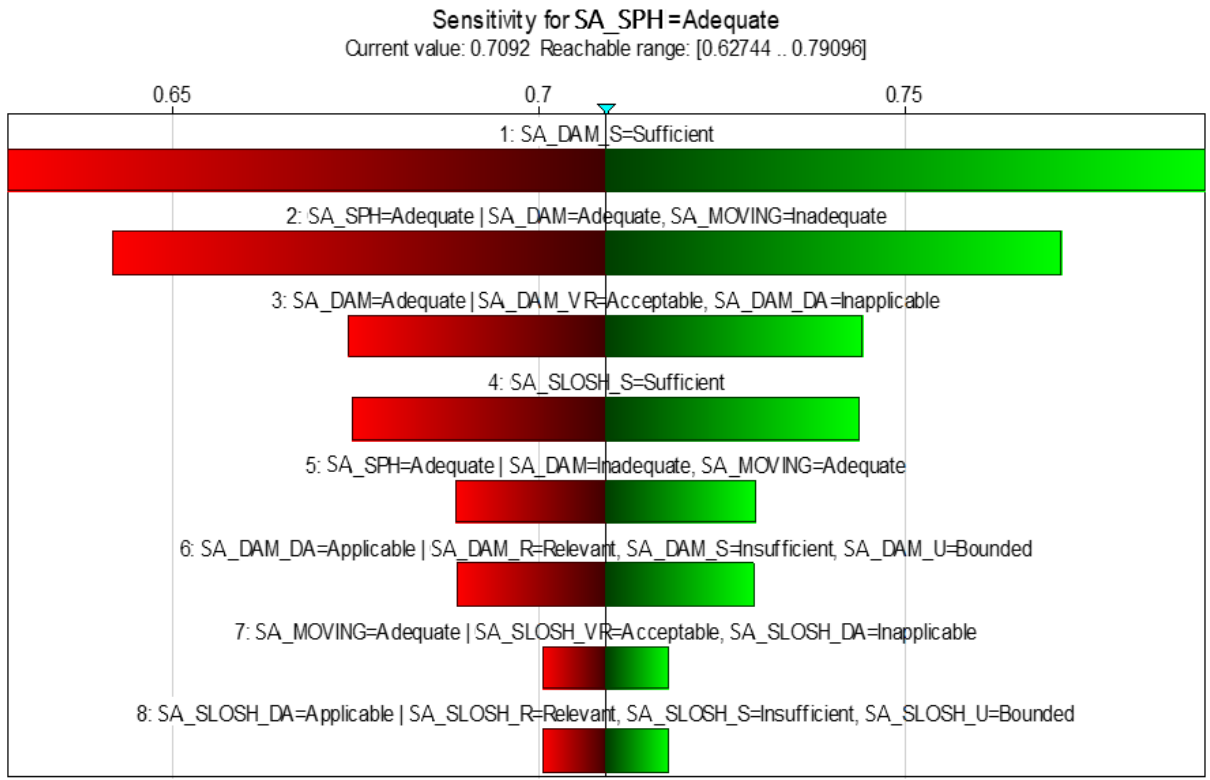

Figure 10: Sensitivity plot for the simulation adequacy assessment with uncertain scaling grade and uncertain conditional probabilities. All conditional probabilities are sampled by 40% of their current values, the maximally reachable belief in an adequate simulation ranges from 0.63 to 0.8.

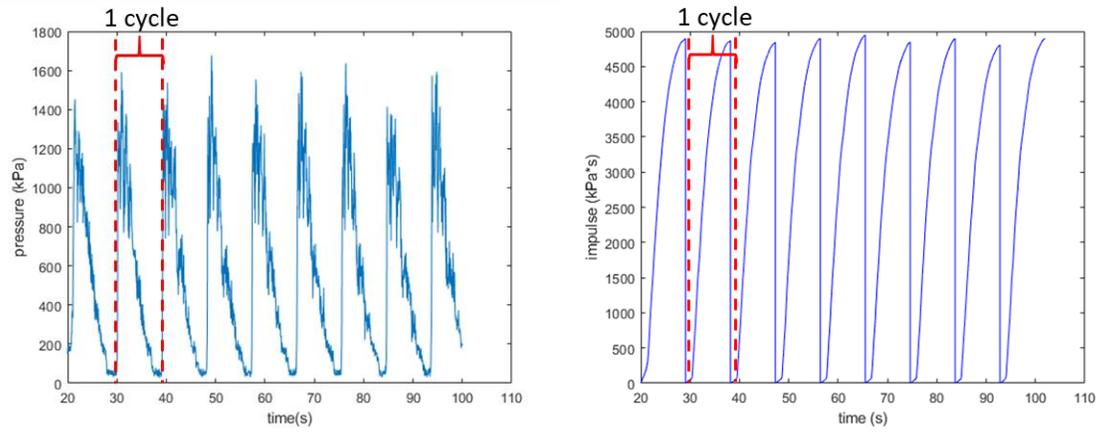

Figure 11: Predicted time transient of hydrodynamic pressures (left) and impulse (right) onto the structure by water surface waves.

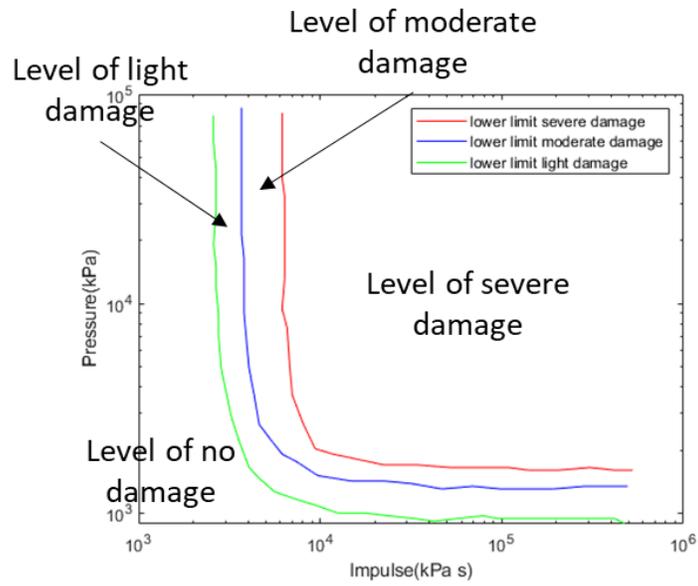

Figure 12: Logarithm plot of damage levels. Four levels of damage are defined based on the P-I values.

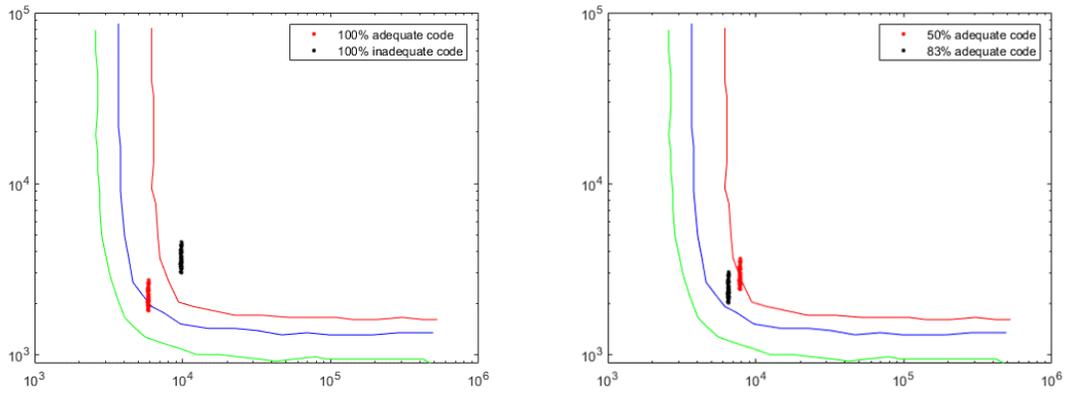

Figure 13: Distribution of P-I values onto the damage levels for all 60 cycles when the simulation is 100% adequate or 100% inadequate (left) and when the simulation is 50% adequate or 83% adequate (right).

Table 1: Important conditions and assumptions with respect to aspects of investigation.

| ID | Conditions/Assumptions |
|----|------------------------|
| **A** | **Investigation Scope** |
| A1 | Current work only focuses on validation activities to support the risk-informed safety analysis |
| A2 | There are data gaps between the validation databases and target applications |
| A3 | The SPH-based computer code is assumed to be verified |
| **B** | **Validation Formalization** |
| B1 | Formalism can reduce the uncertainty |
| B2 | Decision makings in validation is a structured argument process supported by a body of evidence; |
| B3 | Risks due to model uncertainty are characterized by expected losses |
| **C** | **Case Study** |
| C1 | If simulation is not adequate, it can predict the QoIs with maximum 100% errors |
| C2 | If simulation is adequate, it can predict the QoIs with maximum 20% errors |
| C3 | If simulation adequacy is uncertain, QoIs are assembled according to predictions and beliefs at each adequacy level |
| C4 | Preference over different simulation-adequacy levels can be characterized by the magnitude of expected loss |

Table 2:  Assignment of screening probability with characteristics and examples.

| Belief scales in Probabilities | Characteristics | Examples | |
|---|---|---|---|
| | | Validation Result | Data Applicability |
| 1 | Corresponding levels are well-known and obtainable on the major spectrum | Applying CFD-based M&S with very fine grids (DNS scales) | High-quality validation databases are collected from prototypical systems for the directly relevant phenomena |
| 0.1 | Corresponding levels are known but obtainable only at the edge of spectrum | Applying CFD-based M&S with coarser grids (Asymptotic range or outside) | Validation databases are collected from reduced-scale systems for the highly relevant phenomena |
| 0.01 | Corresponding levels cannot be excluded, but it is outside the spectrum of reason | Applying CFD-like or correlation-based M&S | Validation databases are collected from low-quality and reduced-scale systems with the poorly relevant phenomena |
| 0.001 | Corresponding levels are unreasonable and violates well-known reality. Its occurrence can be argued against positively | Applying solid mechanistic M&S | Validation databases are collected from low-quality and reduced-scale systems with irrelevant phenomena |

Table 3: Elements of Toulmin's Argument model with simulation adequacy example.

| | | |
|---|---|---|
| **Claims** | The statement we wish to justify | e.g., Simulation predictive capability for an intended reactor application |
| **Data** | The fact we appeal to, the grounds or information on which our claim is based | e.g., Validation data and results collected from experiments, observations, and knowledge |
| **Warrant** | A statement authorizing the step from data to claim is true; an inference rule | e.g., Results from the scaling analysis that infer system behaviors in prototypical conditions based on validation data in reduced-scale conditions |
| **Backing** | A reason for trusting the warrant | e.g., Argument that authorizes the relevance of investigated phenomena and processes for the target applications |
| **Qualifier** | A term or phrase reflecting the degree to which the data support the claims, e.g. generally, probably | e.g., Argument that evaluates the uncertainty of data and experiment |
| **Rebuttal** | Specific circumstances in which the argument will fail to support the claims as exceptions | e.g., Assumptions and conditions about validation data, model, and adequacy assessment |

Table 4: Example of conditional probabilities based on expert knowledge on causal relationships and dependencies among different evidence characterizations.

| Conditional Probability | Value |
|---|---|
| P(Yes-Applicable data \| No-relevant data) | 0 |
| P(Yes-Applicable data \| Yes-relevant & Yes-scaling & Yes-uncertainty) | 1 |
| P(Yes-Applicable data \| Yes-relevant & Yes-scaling & No-uncertainty) | 0.6 |
| P(Yes-Applicable data \| Yes-relevant & No-scaling & Yes-uncertainty) | 0.2 |
| P(Yes-Applicable data \| Yes-relevant & No-scaling & No-uncertainty) | 0.05 |
| P(Yes-Adequate simulation \| Yes-Applicable data & Yes-Validation result) | 1 |
| P(Yes-Adequate simulation \| Yes-Applicable data & No-Validation result) | 0.3 |
| P(Yes-Adequate simulation \| No-Applicable data & Yes-Validation result) | 0.25 |

Table 5: Summary of parameters in the validation cubic decision model.

| Experimental Measurement Uncertainty (EMU) for a given experiment $J$ | |
|---|---|
| [0, 1] | EMU = 0.001 ⇔ Level 0: Experimental uncertainties are unknown or largely biased |
| | EMU = 0.01 ⇔ Leve 1: Experimental uncertainties are qualitatively analyzed only |
| | EMU = 0.1 ⇔ Level 2: Experimental uncertainties are well characterized for most important measurements, but some remains poorly known |
| | EMU = 1 ⇔ Level 3: Experimental uncertainties are well characterized for all tests |
| **Significance factor for a given experiment $J$: $m$** | |
| [0, 1] | Value ranging from 0 to 1. 0 represents insignificant experiments due to low relevance or low quality. 1 means highly significant experiments that are directly relevant to the applications, and the experimental quality is great. |
| **Significance factor for a given physics $K$: $n$** | |
| [0, 1] | Value ranging from 0 to 1. 0 represents low-ranked phenomena and physics, while 1 means high-ranked ones based on PIRT results |
| **Governing scaling parameters for model $[Sc_{Mod_K}]_{EXP}$ and application $[Sc_{Mod_K}]_{APP}$** | |
| $[0, \infty)$ | Dimensionless quantities that measure the system invariance according to the model |

Table 6: Definition of phases of development.

| Phase # | | Sources of Uncertainty | Levels of Uncertainty |
|---|---|---|---|
| 1 | Scoping | Largely uncertain conditional dependency with unknown bounds. Insufficient evidence or imprecise beliefs on evidence with uncertain bounds. Unverified or low-quality evidence. | Uncertainty in the final simulation adequacy is so large that preliminary sensitivity analysis shows that the uncertainty will alter the decisions in designated scenarios and applications. |
| 2 | Refinement | Uncertain conditional dependency with known bounds. Sufficient evidence with imprecise beliefs but known bounds. | Uncertainty in the final simulation adequacy has known ranges or distributions with confidence levels. The uncertainty can alter the decisions only at the edge of scenario spectrum |
| 3 | Maturation | Conditional dependency with precisely known distributions. Sufficient evidence with beliefs on evidence and precisely known distributions. | Uncertainty in the final simulation adequacy is precisely characterized, and they are not likely to alter the decisions. |

Table 7. Validation results for SPH methods in simulating hydrodynamic forces on stationary structures in the external-flooding scenario.

| Benchmark | Simulation Adequacy | Accuracy (L1 error) | Data Applicability | | |
|---|---|---|---|---|---|
| | | | Relevancy | Scaling | Data Quality |
| Dam Break | Adequate | Acceptable ($L_1$=3.6%) | Yes | Yes | High |
| Moving Solids in Fluid | Inadequate | Falling: Acceptable ($L_1$=5.52%) | Yes | No | High |
| | | Floating: Acceptable ($L_1$=4.41%) | | | |

Table 8: A list of decision parameters in validation cubic model. The value is assigned based on author's knowledge.

| Decision Parameters | Dam Break | Moving Solid in Fluid |
|---|---|---|
| $m$ | 1 (Low) | 3 (High) |
| $n$ | 3 (High) | 3 (High) |
| $\left[Sc_{Mod\_K}\right]_{EXP}$ | 0.1~0.26 ($x^*$ EXP) | 0.017 ($x^*$ EXP) |
| $\left[Sc_{Mod\_K}\right]_{APP}$ | 0.1 ($x^*$ APP) | 0.1 ($x^*$ APP) |
| $RPP$ | 1 | 0.17 |
| $EMU$ | 0.1 (Level 1) | 0.01 (Level 2) |
| $\psi_{E_i}$ | 0.43 | 0.16 |
| $\overline{\psi_{E_t}}$ | 0.73 | 0.27 |

Table 9: Number of cycles in each damage levels for four different simulation adequacy results.

| Damage Level (DL) | Loss (C) | Probability of Occurrence $P_{DL}$ among 60 cycles | | | |
|---|---|---|---|---|---|
| | | 100% Adequate | 100% Inadequacy | Ensembled 50/50 | Ensembled 83/17 |
| *No* | 0 | 0/60 | 0/60 | 0/60 | 0/60 |
| *Light* | $10 | 21/60 | 0/60 | 0/60 | 0/60 |
| *Moderate* | $50 | 39/60 | 0/60 | 34/60 | 60/60 |
| *Severe* | $100 | 0/60 | 60/60 | 26/60 | 0/60 |

Table 10: Expected losses for four simulation adequacy results.

| Decision Analysis | Expected Loss $\langle C \rangle$ |
|---|---|
| **Qualitative and Implicit framework (Ensembled 50/50)** | $71.67 |
| **PCMQBN decision framework (Ensembled 83/17)** | $50 |
| **Optimistic decision maker (100% Adequate)** | $36 |
| **Conservative decision maker (100% Inadequate)** | $100 |